\documentclass[preprint,12pt]{elsarticle}

\usepackage{amsmath}
\usepackage{amssymb}
\usepackage{amsthm}
\usepackage{latexsym} 
\usepackage{xcolor} 

\usepackage{graphicx}
\usepackage{booktabs}
\usepackage{multirow}
\usepackage{diagbox}
\usepackage{adjustbox}
\usepackage{array}
\usepackage{makecell}
\usepackage{stfloats}
\usepackage{framed}

\usepackage[font=footnotesize,labelformat=simple]{subcaption}

\usepackage{url}
\usepackage{bbding}
\usepackage{geometry}
\usepackage{tabularx}

\usepackage{hyperref}      

\usepackage{lineno}

\usepackage[most]{tcolorbox}
\newenvironment{hlbox}{}{}

\usepackage{soul}
\renewcommand{\hl}[1]{#1} 

\renewcommand{\colorbox}[2]{#2}

\journal{Medical Image Analysis}

\begin{document}

\begin{frontmatter}


\begin{graphicalabstract}
\centering
\includegraphics[width=\linewidth]{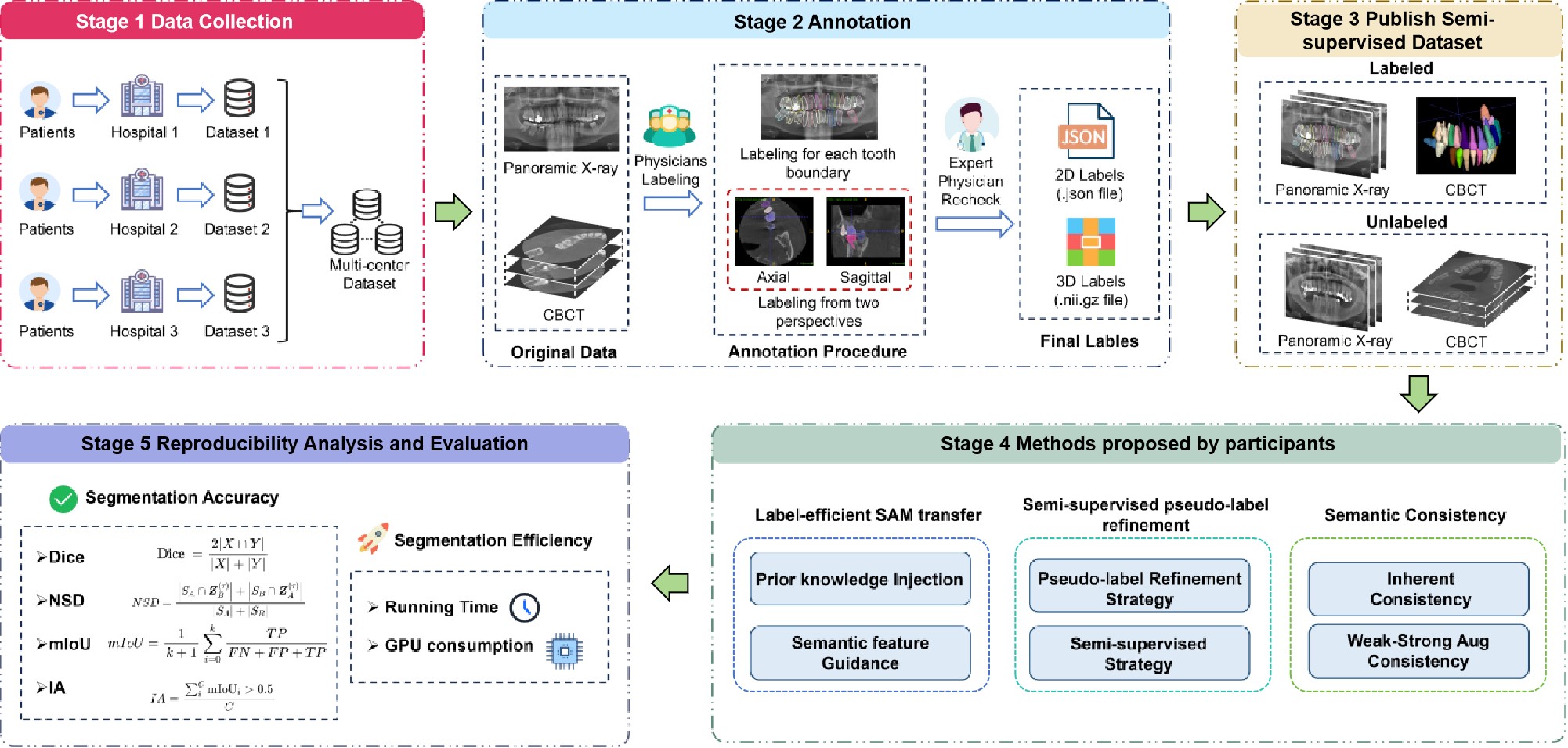}
\par\addvspace{10pt} 
\small 
End-to-end workflow of the MICCAI 2024 Semi-supervised Teeth Segmentation (STS) Challenge. The process encompasses five key stages: (1) multi-center data collection to ensure dataset diversity, (2) iterative annotation by clinicians for high-quality ground truth, (3) construction of the semi-supervised dataset with distinct labeled and unlabeled sets, and (4/5) the final summarization and evaluation of submitted participant methods.
\end{graphicalabstract}

\begin{highlights}
    \item A novel dataset for semi-supervised instance-level tooth segmentation. 
    \item First international benchmark of SSL for tooth instance segmentation. 
    \item SSL boosts instance segmentation accuracy by over 60 percentage points compared to a supervised baseline. 
    \item Analysis of top methods including SAM integration and pseudo-labeling.
    \item Provides insights for label-efficient AI in clinical dental imaging.
\end{highlights}

\title{MICCAI STS 2024 Challenge: Semi-Supervised Instance-Level Tooth Segmentation in Panoramic X-ray and CBCT Images}


\cortext[cor1]{Corresponding authors: ljun77@hdu.edu.cn (Jun Liu); shuaiwang.tai@gmail.com (Shuai Wang), hz143@leicester.ac.uk (Huiyu Zhou)}

\fntext[eq]{These authors contributed equally to this work.}

\author[1]{Yaqi {Wang}\fnref{eq}}
\author[4]{Zhi {Li}\fnref{eq}} 
\author[12]{Chengyu {Wu}}
\author[1]{Jun {Liu}\corref{cor1}}
\author[3]{Yifan {Zhang}}

\author[5]{Jiaxue {Ni}}
\author[5]{Qian {Luo}}
\author[5]{Jialuo {Chen}}
\author[10]{Hongyuan {Zhang}}
\author[5]{Jin {Liu}}
\author[18]{Can {Han}}

\author[17]{Kaiwen {Fu}}
\author[18]{Changkai {Ji}}
\author[19]{Xinxu {Cai}}
\author[20]{Jing {Hao}}
\author[25]{Zhihao {Zheng}}
\author[26]{Shi {Xu}}
\author[27]{Junqiang {Chen}}

\author[11]{Qianni {Zhang}}
\author[18]{Dahong {Qian}}
\author[4]{Shuai {Wang}\corref{cor1}}
\author[110]{Huiyu {Zhou}\corref{cor1}}


\address[1]{Innovation Center for Electronic Design Automation Technology, Hangzhou Dianzi University, Hangzhou, China}
\address[4]{School of Cyberspace, Hangzhou Dianzi University, Hangzhou, China}
\address[5]{Hangzhou Dianzi University, Hangzhou, China}

\address[3]{Hangzhou Geriatric Stomatology Hospital, Hangzhou Dental Hospital Group, Hangzhou, China}
\address[10]{Shenzhen University, Shenzhen, China}
\address[11]{Queen Mary University of London, London, United Kingdom}
\address[12]{Shandong University, Weihai, China}
\address[17]{Xidian University, Xi’an, China}
\address[18]{Shanghai Jiao Tong University, Shanghai, China}
\address[19]{Harbin Institute of Technology, Harbin, China}
\address[20]{The University of Hong Kong, Hong Kong SAR, China}
\address[25]{Chinese Academy of Sciences, Chengdu, China}
\address[26]{Yunnan Provincial Stomatology Hospital, Kunming, China}
\address[27]{Shanghai MediWorks Precision Instruments Co., Ltd, Shanghai, China}
\address[110]{University of Leicester, Leicester, United Kingdom}

\begin{abstract}
Orthopantomogram (OPGs) and Cone-Beam Computed Tomography (CBCT) are vital for dentistry, but creating large datasets for automated tooth segmentation is hindered by the labor-intensive process of manual instance-level annotation. This research aimed to benchmark and advance semi-supervised learning (SSL) as a solution for this data scarcity problem. We organized the 2nd Semi-supervised Teeth Segmentation (STS 2024) Challenge at MICCAI 2024.
We provided a large-scale dataset comprising over 90,000 2D images and 3D axial slices, which includes 2,380 OPG images and 330 CBCT scans, all featuring detailed instance-level FDI annotations on part of the data. The challenge attracted 114 (OPG) and 106 (CBCT) registered teams. \hl{To ensure algorithmic excellence and full transparency, we rigorously evaluated the valid, open-source submissions from the top 10 (OPG) and top 5 (CBCT) teams, respectively.} All successful submissions were deep learning-based SSL methods. The winning semi-supervised models demonstrated impressive performance gains over a fully-supervised nnU-Net baseline trained only on the labeled data. For the 2D OPG track, the top method improved the Instance Affinity (IA) score by over 44 percentage points. For the 3D CBCT track, the winning approach boosted the Instance Dice score by 61 percentage points. This challenge confirms the substantial benefit of SSL for complex, instance-level medical image segmentation tasks where labeled data is scarce. 
The most effective approaches consistently leveraged hybrid semi-supervised frameworks that combined knowledge from foundational models like SAM with multi-stage, coarse-to-fine refinement pipelines. \hl{Both the challenge dataset and the participants’ submitted code have been made publicly available on GitHub }(\url{https://github.com/ricoleehduu/STS-Challenge-2024}), ensuring transparency and reproducibility.
\end{abstract}


\end{frontmatter}



\section{Introduction}
\begin{figure*}[!ht]
\centering
\includegraphics[width=\linewidth]{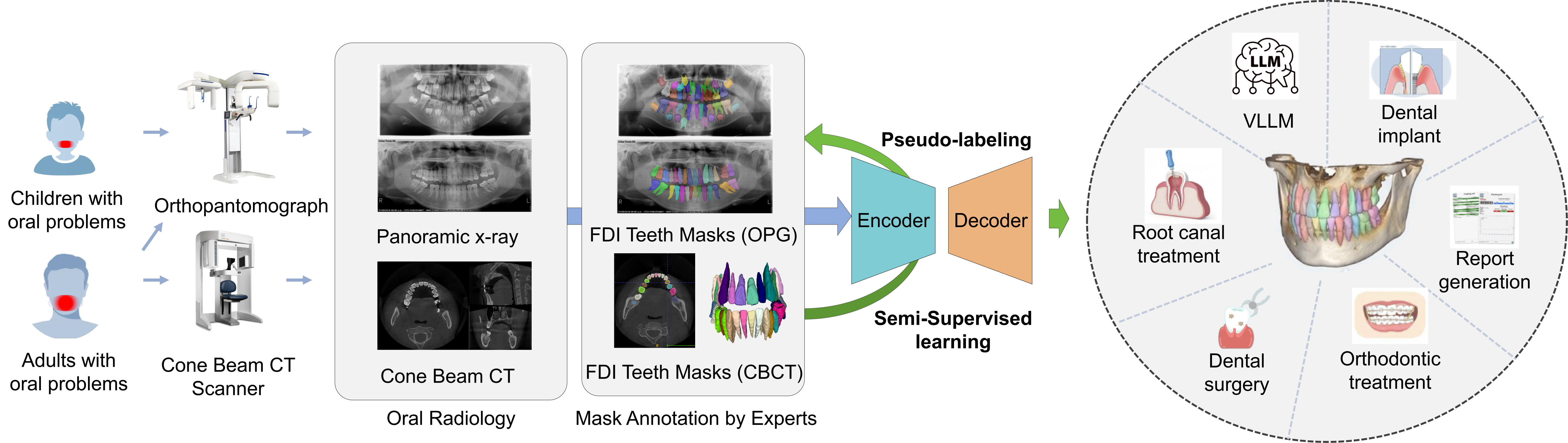}
\caption{Overview of the Semi-Supervised Learning (SSL) framework for dental instance segmentation and its clinical utility. The workflow proceeds from (Left) multi-modal data acquisition (OPG and CBCT) covering both pediatric and adult populations, to (Middle) the core semi-supervised training paradigm where an Encoder-Decoder network leverages expert annotations and iteratively refines performance via pseudo-labeling, and finally to (Right) diverse downstream clinical applications, such as orthodontic treatment planning, root canal therapy, and automated report generation, which rely on precise tooth instance masks. }
\label{fig:fig1}
\end{figure*}

Oral health is integral to overall well-being, with dental structures profoundly impacting an individual's quality of life~\cite{sischo2011oral}. Modern dentistry relies heavily on advanced imaging modalities such as 2D Orthopantomogram (OPGs) and 3D Cone-Beam Computed Tomography (CBCT)~\cite{shah2014recent}. OPGs offer a broad view of the dentition and surrounding structures, while CBCT provides high-resolution 3D data for detailed anatomical evaluation, crucial for detecting caries, impacted or supernumerary teeth, and for precise diagnosis and treatment planning in orthodontics and implantology~\cite{cosson2020interpreting,jain2019new}. Accurate tooth segmentation from these images is fundamental for many computer-aided systems, enhancing clinical decision making, particularly when performed at the instance level~\cite{xu20183d}. Therefore, robust automated instance-level tooth segmentation algorithms are essential to advance the precision, efficiency, and personalization of dental care~\cite{lahoud2021artificial}, a process visualized in Fig. \ref{fig:fig1}.

Despite its clinical importance, the development of high-performance automated tooth segmentation models is critically hampered by data scarcity. Manual annotation, especially for detailed instance-level identification, is extremely time-consuming, laborious, and requires expert knowledge~\cite {sapkota2024zero}. This results in a critical shortage of large-scale, high-quality labeled datasets essential for training conventional fully supervised deep learning models. Consequently, while fully supervised methods show promise, their performance is often limited by the lack of diverse and comprehensively annotated data, especially for complex cases.

Semi-supervised learning (SSL) has emerged as a promising paradigm to address limited labeled data in medical image analysis, including dental imaging~\cite{redmon2016you}. SSL methods leverage a small labeled dataset alongside a larger pool of unlabeled images, reducing reliance on extensive manual annotation and potentially improving model generalization and robustness. However, SSL presents its own challenges. A core difficulty is generating and refining high-quality pseudo-labels for unlabeled data, as inaccuracies can propagate errors and degrade performance. Models may also suffer from confirmation bias. Thus, effective SSL requires careful design of consistency regularization, uncertainty estimation, and pseudo-label selection mechanisms to ensure genuine learning from unlabeled data without over-reliance on the initial labeled set~\cite{qin2023flexssl}.

Tooth segmentation becomes considerably more complex when progressing from semantic delineation (tooth vs. background) to instance segmentation. Instance-level segmentation demands not only precise boundary delineation but also unique identification of each tooth, often with its FDI number, effectively treating each tooth instance as a distinct class. This complexity is amplified by variable dental anatomy, such as morphology, arrangement, crowding, and impaction, across diverse age groups and pathologies. Close tooth proximity and overlap, especially in 2D OPGs, hinder accurate instance separation. Image quality issues in OPG and CBCT, like distortions, artifacts, and varying contrast, add further difficulty~\cite {kumarihami2018development}. Moreover, integrating robust FDI numbering within an SSL framework with sparse ground truth for such identification is a significant hurdle, compounded by data imbalances and the need for generalization across varied clinical settings and acquisition protocols.

\begin{table}[ht]
\centering
\caption{Summary of labeled and unlabeled data statistics for the 2D dental panoramic X-ray and 3D Cone-Beam Computed Tomography (CBCT) datasets, including patient/scans counts, image resolution, and anatomical annotations.}
\label{tab:dataset_statistics}
\setlength{\tabcolsep}{5mm}
\renewcommand\arraystretch{1.3}
\resizebox{\linewidth}{!}{
\begin{tabular}{llcc}
\toprule
\textbf{Dataset} & \textbf{Metric} & \textbf{Labeled} & \textbf{Unlabeled} \\
\midrule
\multirow{7}{*}{OPG(2D)}
    & Child Samples           & 30     & 887 \\
    & Adult Samples           & 70     & 1484 \\
    & Number of Patients      & 100    & 2323 \\

    & Resolutions (Child)(Pixel)   & $2000 \times 942$      & $2000 \times 942$ \\
    & Resolutions (Adult)(Pixel)     & $1991 \times 1127$     & $2000 \times 942$ (887) / $1991 \times 1127$ (1463) \\
    & Number of Teeth         & $\approx 2700$   & $\approx 65044$ \\
\midrule
\multirow{7}{*}{CBCT(3D)}
    & Adult Samples           & 100    & 300 \\
    & Number of Patients      & 100    & 300 \\
    & Number of Slices         & 29240& 60000 \\
    & Resolutions(Voxel)             & $266 \times 266$ / $512 \times 512$ & $266 \times 266$ \\
    & In-plane Resolution (mm)& $0.3 \times 0.3$ / $0.25 \times 0.25$ & $0.3 \times 0.3$ \\
    & Slice Thickness (mm)    & 0.3 / 0.25 & 0.3 \\
    & Number of Teeth         & $\approx 3000$   & $\approx 9000$ \\
\bottomrule
\end{tabular}
}
\end{table}

Despite the recognized need for automated dental image analysis, developing and benchmarking advanced algorithms, especially for instance-level tasks, is hindered by limited public datasets. Although there are several dental datasets~\cite{jiang2023instance}, many have limitations for robust semi-supervised instance-level systems (Table~\ref{tab:dental_datasets}, \ref{tab:dental_cbct_datasets_updated}). For example, some focus on adult populations or lack granular FDI-numbered annotations crucial for clinical applications. Critically, most are not designed for SSL, often missing large, matched, unlabeled image sets. Recognizing these gaps, we initiated the Semi-supervised Teeth Segmentation (STS) challenge series. Although our inaugural STS 2023 challenge established a baseline for semantic segmentation~\cite{wang2025miccai}, the more clinically crucial and complex task of instance-level segmentation remained an open problem, motivating the current work.

\begin{figure*}[!ht]
\centering
\includegraphics[width=\linewidth]{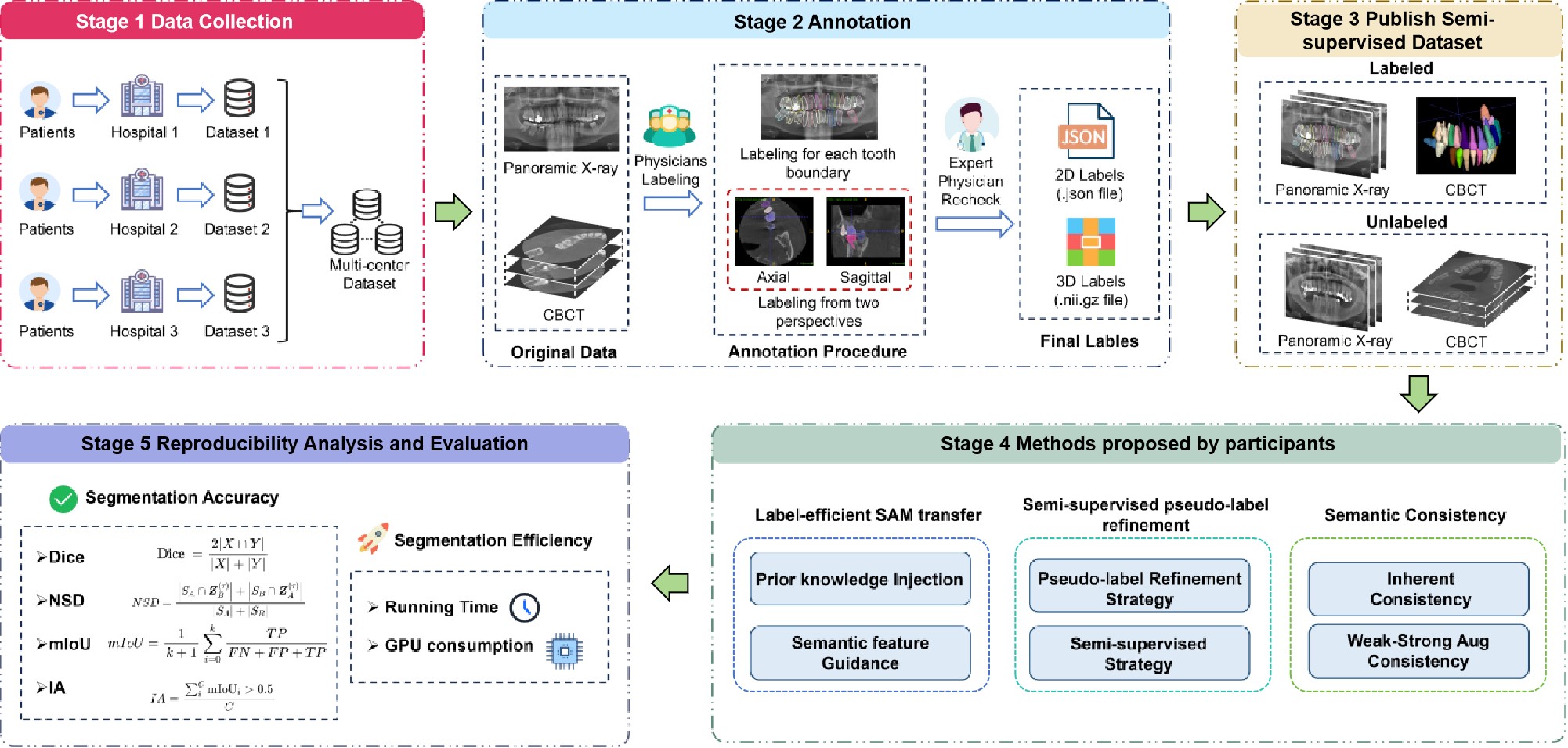}
\caption{End-to-end workflow of the MICCAI 2024 Semi-supervised Teeth Segmentation (STS) Challenge. The process encompasses five key stages: (1) multi-center data collection to ensure dataset diversity, (2) iterative annotation by clinicians for high-quality ground truth, (3) construction of the semi-supervised dataset with distinct labeled and unlabeled sets, and (4/5) the final summarization and evaluation of submitted participant methods.}
\label{fig:framework}
\end{figure*}

Building on this foundation, the 2nd Semi-supervised Teeth Segmentation (STS 2024) Challenge was organized as an official satellite event of the 27th International Conference on Medical Image Computing and Computer-Assisted Intervention (MICCAI 2024). Its primary aim was to stimulate the development and rigorous evaluation of novel SSL algorithms for instance-level tooth segmentation in both 2D OPGs and 3D CBCTs, creating a standardized benchmark for this challenging task. The challenge featured two tasks (OPG and CBCT), requiring semi-supervised instance segmentation and identification of up to 32 permanent and 20 deciduous teeth. Participants received datasets with a small fraction of instance-level annotations and a majority of unlabeled data as described in Table~\ref{tab:dataset_statistics}, with evaluations covering both accuracy and efficiency. \hl{The challenge dataset, including the labeled training set, validation set , and the large-scale unlabeled set, is now publicly available for direct download  in our GitHub repository} (\url{https://github.com/ricoleehduu/STS-Challenge-2024}). The complete workflow of the STS 2024 Challenge is shown in Fig. \ref{fig:framework}.

In general, the main contributions are summarized as follows:
\begin{itemize}
    \item A novel public dataset for semi-supervised instance-level tooth segmentation, the first large-scale, multi-modal resource including OPG and CBCT, specifically curated for semi-supervised instance-level tooth segmentation. The dataset features detailed FDI annotations for a small labeled set alongside extensive unlabeled data, covering diverse patient demographics and pathologies.
    \item We established the first open international benchmark for this specific task, employing a rigorous evaluation framework with Dockerized submissions on a hidden test set to ensure fair and reproducible comparisons of state-of-the-art methods.
    \item  Provide a detailed analysis and summary of the various SSL strategies employed by the top-performing participants. This includes an overview of popular architectures, effective SSL techniques, and data handling approaches, offering valuable insights to guide future research in label-efficient AI for dentistry.
\end{itemize}

\begin{table*}[ht]
\centering
\caption{Comparison of the STS 2024 dataset with other public 2D OPG datasets. The STS 2024 dataset provides a unique resource for semi-supervised learning, featuring a large unlabeled set, multi-center data, and detailed instance-level annotations across a wide age range. "A / woA" denotes data with/without annotations.}
\label{tab:dental_datasets}
\setlength{\tabcolsep}{2.5mm} 
\renewcommand\arraystretch{1.3}  
\resizebox{\linewidth}{!}{ 
\begin{tabular}{@{}l l c c l c c l@{}} 
\toprule
\textbf{Dataset} & \textbf{Year} & \makecell[c]{\textbf{X-Ray Slices} \\ \textbf{(A / woA)}} & \makecell[c]{\textbf{Num. of} \\ \textbf{Teeth}} & \makecell[c]{\textbf{Annotation } \\ \textbf{Type}} & \textbf{Annotators} & \makecell[c]{\textbf{Centers}} & \makecell[c]{\textbf{Resolution} \\ \textbf{(Pixels)}} \\
\midrule
\makecell[l]{Dental Panoramic\\X-Rays \cite{abdi2017panoramic}} & 2017 & \makecell[c]{116 \\ (116 / 0)} & $\approx$ 3,480 & Tooth Instances & 2 & 1 & $3100\times1300$ \\
UFBA-UESC \cite{budagam2024instance} & 2019 & \makecell[c]{1,500 \\ (543 / 957)} & $\approx$ 45,000 & \makecell[l]{Instance, \\ Numbering} & 1 & 1 & $1991\times1127$ \\
Tufts Dental \cite{panetta2021tufts} & 2021 & \makecell[c]{1,000 \\ (1000 / 0)} & $\approx$ 27,000 & \makecell[l]{Semantic, \\ Caries Attn.} & 2 & 1 & $1615\times840$ \\
\makecell[l]{Panoramic\\Radiography \cite{roman2021panoramic}} & 2021 & \makecell[c]{598 \\ (0 / 598)} & $\approx$ 17,940 & \makecell[l]{Lesion \\ Segmentation} & 1 & 1 & $2041\times1024$ \\
DENTEX \cite{hamamci2023dentexabnormaltoothdetection} & 2023 & \makecell[c]{2,332 \\ (1005 / 1227)} & $\approx$ 69,960 & \makecell[l]{Tooth Bounding\\Box} & 3 & 1 & $2098\times970$ \\
vzrad2 \cite{vzrad2_dataset} & 2024 & \makecell[c]{8,188 \\ (8188 / 0)} & $\approx$ 245,640 & \makecell[l]{Pathology, \\ Treatment} & 1 & 1 & $640\times640$ \\
STS2023 \cite{wang2025miccai} & 2023 & \makecell[c]{6,500 \\ (6500 / 0)} & $\approx$ 195,000 & Semantic & 30 & 2 & $640\times640$ \\
\textbf{STS2024 (Ours)} & \textbf{2024} & \makecell[c]{\textbf{2,380} \\ \textbf{(30 / 2350)}} & \textbf{$\approx$ 71,400} & \makecell[l]{\textbf{Instance,} \\ \textbf{Numbering}} & \textbf{30} & \textbf{2} & \makecell[l]{\textbf{$2000\times942$}\\\textbf{$1991\times1127$}} \\
\bottomrule
\end{tabular}
}
\end{table*}

\begin{table*}[ht]
\centering
\caption{Comparison of STS2024 dataset and other public 3D dental CBCT datasets. The STS2024 dataset provides broader age coverage, more detailed annotations, and multi-center diversity, making it a valuable resource for clinical dental AI research.}
\label{tab:dental_cbct_datasets_updated}
\setlength{\tabcolsep}{3mm} 
\renewcommand\arraystretch{1.3}
\resizebox{\linewidth}{!}{ 
\begin{tabular}{lllllllcl}  %
\toprule
\textbf{Dataset} & \textbf{Years} & \makecell[c]{\textbf{Age} \\ \textbf{Range}} & \textbf{Patients} & \makecell[c]{\textbf{Volumes} \\ \textbf{(A / woA)}} & \textbf{Slices} & \makecell[c]{\textbf{Num of} \\ \textbf{Teeth}} & \makecell[c]{\textbf{Annotator}} & \makecell[c]{\textbf{Volumes} \\ \textbf{(pixels)}} \\
\midrule
\makecell[l]{Clinical dental \\ CBCT~\cite{9083982}}
           & 2020 & 10-49      & 25  & \makecell[c]{25 \\ (25 / 0)}             & 9400   & $\approx$ 770   & 1  & \makecell[c]{$110\times145\times280$} \\
CTooth+\cite{cui2022ctooth+}                       & 2022 & 10--15 & 22  & \makecell[c]{168 \\ (22 / 146)} & 31380  & $\approx$ 5040  & 15  & \makecell[c]{$266\times266\times266$} \\
\makecell[l]{Mandibular Canal \\ CBCT~\cite{9686728}}
  & 2022 & 10-100      & 347   & \makecell[c]{347 \\ (347 / 0)}           & 88832      & $\approx$ 17220 & -  & \makecell[c]{$178\times423\times463$} \\
\makecell[l]{Multi-modal \\ dataset~\cite{huang2024multimodal}}
             & 2023 & 18-89      & 169   & \makecell[c]{188 \\ (188 / 188)} & 16203    & $\approx$ 5401  & 13 & \makecell[c]{$640\times640\times200$} \\
STS2023~\cite{wang2025miccai} & 2023 & 7--70  & 584 & \makecell[c]{584 \\ (84 / 500)}& 88500  & $\approx$ 17520  & 30 & \makecell[c]{$640\times640\times399$} \\
\hl{ToothFairy}\cite{lumetti2024enhancing} & 2022 & 12-100 & 443 & \makecell[c]{443\\(443/ 0)} & 64000 & -- (IAC only) & 5 & \makecell[c]{$169\times342\times370$} \\
\hl{ToothFairy2}\cite{bolelli2024segmenting} & 2023 & 16-100 & 530 & \makecell[c]{530\\(480/50)} & 80000 & $\approx$ 71400 & 5 & \textbf{\makecell[c]{$170\times272\times345$\\$298\times512\times512$}} \\
\hl{ToothFairy3}\cite{bolelli2025segmenting} & 2024 & 16-100 & 532 & \makecell[c]{532\\(532/0)} & 80000 & $\approx$ 17024 & 5 & \textbf{\makecell[c]{$170\times272\times345$\\$298\times512\times512$}} \\
\textbf{STS2024 (Ours)  }                       & \textbf{2024} & \textbf{7--70}  & \textbf{330} & \makecell[c]{\textbf{330} \\ \textbf{(30 / 300)}} & \textbf{69960}  & \textbf{$\approx$ 9900}  & \textbf{30} & \textbf{\makecell[c]{$266\times266\times200$\\$512\times512\times332$}} \\
\bottomrule
\end{tabular}
}
\end{table*}

\section{Related Works}

\subsection{Datasets and Benchmarks in Dental Image Segmentation}
Artificial intelligence (AI) is increasingly vital in stomatology, where AI-driven segmentation of teeth and maxillofacial structures from Orthopantomogram (OPGs) and Cone-Beam Computed Tomography (CBCT) is a critical enabler for applications like lesion visualization, orthodontic design, and surgical planning~\cite{arsiwala2023machine}. This has led to the creation of numerous public datasets to fuel research in this area (see Tables~\ref{tab:dental_datasets} and \ref{tab:dental_cbct_datasets_updated}). However, a closer examination reveals critical gaps, particularly for developing robust, label-efficient, instance-level segmentation models.

Many existing datasets, while valuable, have significant limitations for our task. For instance, some primarily offer semantic-level annotations, which cannot distinguish between individual teeth. The Tufts Dental~\cite{panetta2021tufts} dataset and our own previous STS2023~\cite{wang2025miccai} dataset fall into this category, providing strong baselines for tooth-background separation but not for instance identification. Other datasets provide instance-level information, but with limitations in scope or annotation type. For example, DENTEX~\cite{hamamci2023dentexabnormaltoothdetection} provides only bounding boxes rather than fine-grained masks, and datasets like UFBA-UESC~\cite{budagam2024instance} and CTooth+~\cite{cui2022ctooth+} focus on specific populations, such as excluding pediatric teeth or having a narrow age range, which limits model generalizability to diverse clinical scenarios.

\hl{While theoretically any fully annotated dataset can be adapted for semi-supervised learning (SSL) by withholding labels, the specific challenge of dental instance segmentation demands a dataset structure that mirrors real-world clinical constraints. The meticulous, pixel-level instance annotation required for tasks like root canal therapy and maxillofacial surgery creates a severe annotation bottleneck. Our STS 2024 dataset addresses this by providing a curated benchmark specifically designed to test SSL algorithms on a difficult multi-class instance segmentation task. It combines high-quality, expert-verified instance masks for a small labeled set with a massive, clinically representative corpus of unlabeled data, directly simulating the scenario where leveraging abundant raw data is the only viable path to high performance.}

To directly address these multifaceted limitations, we initiated the Semi-supervised Teeth Segmentation (STS) challenge series. Although our inaugural STS 2023 challenge~\cite{wang2025miccai} established a robust baseline for SSL-based semantic segmentation, the ability to distinguish and number individual teeth, a prerequisite for most advanced clinical applications, remained an unaddressed challenge. To bridge this critical gap, the STS 2024 Challenge was designed to elevate the task to semi-supervised instance-level segmentation. The challenge introduces a unique, multimodal dataset with detailed, instance-level FDI annotations complemented by a vast pool of unlabeled OPG and CBCT data. Spanning a wide range of ages (7--70 years) and pathologies, this competitive framework is the first of its kind designed to catalyze and rigorously evaluate SSL algorithms for precise, instance-aware segmentation, with the aim of advancing intelligent and efficient dental care.

\subsection{Methodologies for Tooth Segmentation and Semi-Supervised Learning}
Automated tooth segmentation has progressed from traditional image processing to deep learning (DL). In fully supervised settings, U-Net~\cite{ronneberger2015u} and its variants like V-Net~\cite{milletari2016v} and nnU-Net~\cite{isensee2021nnu} are standard for semantic segmentation in dental imaging~\cite {dot2024dentalsegmentator, wang2024trans}. For instance-level segmentation, which distinguishes individual teeth, common approaches adapt computer vision methods like Mask R-CNN~\cite{he2017mask} or use post-processing techniques to separate instances after semantic segmentation. However, the performance of these fully supervised methods is fundamentally tethered to the availability of large, meticulously annotated datasets, a condition rarely met in specialized medical domains like dentistry.

Semi-supervised learning (SSL) mitigates this dependency on labeled data. One prominent SSL paradigm is pseudo-labeling, also known as self-training. In this approach, a model first trains on a small labeled set and then generates predictions, or pseudo-labels, for unlabeled data. These new labels are then used to augment the training set, allowing the model to learn from the unlabeled images~\cite{lee2013pseudo}. A critical challenge here is managing the quality of these pseudo-labels to prevent error propagation. This is often addressed through techniques like confidence thresholding and iterative refinement~\cite{sohn2020fixmatch}. However, in dense dental anatomies, generating high-confidence pseudo-labels for heavily overlapping or small, developing teeth remains a significant challenge.
Another major SSL family is consistency regularization. The core principle is that a model's prediction for an unlabeled input should remain consistent even when the input is perturbed, for example, through data augmentation or internal model changes like dropout~\cite{tarvainen2017mean}. The Mean Teacher model is a classic example of this, where a "student" network is trained to produce predictions consistent with a "teacher" network, which is an exponential moving average of the student's own weights~\cite{tarvainen2017mean}. A key challenge for consistency-based methods in dental imaging is defining realistic augmentations for OPG/CBCT data without altering the fine anatomical boundaries critical for segmentation.

Many state-of-the-art SSL methods combine these two strategies. They often use teacher-student frameworks to guide learning on unlabeled data via robust pseudo-label generation~\cite{berthelot2019mixmatch, chen2020simple}. Other techniques like multi-stage training and uncertainty-aware learning are also employed to enhance performance~\cite{wang2023multi}. While SSL has been successful in various medical segmentation tasks~\cite{cheplygina2019not, tajbakhsh2020embracing}, its application to instance-level dental segmentation was relatively limited before initiatives like the STS challenges~\cite{wang2023multi}.

Emerging DL trends are also shaping this field. Foundation models, such as the Segment Anything Model (SAM)~\cite{kirillov2023segment}, provide powerful pre-trained capabilities that can be used for few-shot segmentation or to generate high-quality initial pseudo-labels, potentially overcoming the difficulty of separating overlapping teeth where traditional models fail~\cite{cheng2023sam}. Another effective strategy is self-supervised pre-training on large unlabeled datasets before fine-tuning with SSL~\cite{chen2020simple}. Finally, automated frameworks like nnU-Net~\cite{isensee2021nnu} serve as strong baselines that can be adapted for SSL. The STS 2024 challenge, therefore, provides a timely and crucial benchmark to evaluate how these diverse SSL strategies and emerging techniques contend with the real-world complexities of instance-level tooth segmentation.

\begin{figure*}[!ht]
\centering
\includegraphics[width=\linewidth]{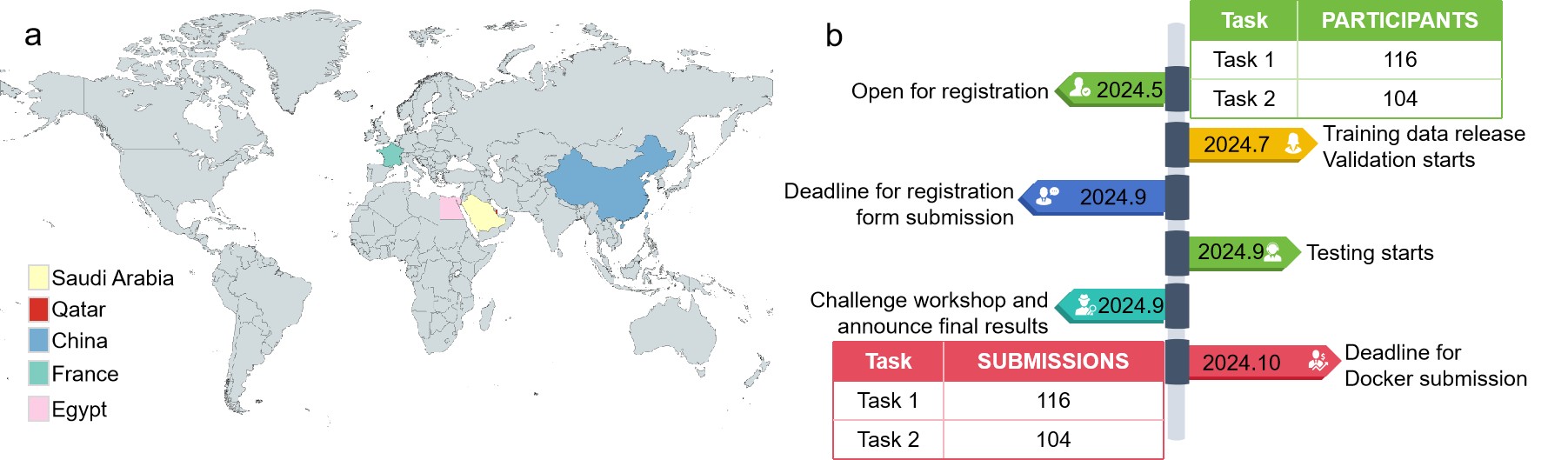}
\caption{Overview of the STS 2024 Challenge participation and timeline. (a) A world map illustrating the geographical distribution of registered participants. (b) A detailed timeline of the challenge schedule, from the training phase start to the final announcement of results.}
\label{fig:schedule}
\end{figure*}

\section{Challenge Description}
\subsection{Dataset information and annotation details}
The STS (2nd Semi-Supervised Tooth Segmentation) Challenge is one of the official challenges of MICCAI 2024, which aims to advance SSL-based tooth image segmentation, focusing on algorithmic performance in multi-instance, multi-class scenarios. The challenge required participants to leverage a small labeled dataset and a large unlabeled dataset to achieve accurate, instance-level tooth segmentation and classification. The challenge consists of two tasks: Task 1 is based on 2D panoramic radiographs (OPG) and Task 2 is based on 3D cone beam CT (CBCT), where up to 32 permanent teeth and 20 deciduous teeth need to be accurately recognized, reflecting the full spectrum of possible dentition. The challenge was hosted on the Codabench platform and attracted global participation, as illustrated by the participant distribution map and competition timeline in Fig. \ref{fig:schedule}. The schedule was divided into distinct phases, including training, validation, and a final testing phase on a hidden dataset to ensure a fair and rigorous evaluation of all submitted methods.

The datasets used in this challenge were provided by Hangzhou Stomatological Hospital and Hangzhou Qiantang Stomatological Hospital, and were from real clinical environments, covering a wide range of pre-treatment images such as missing teeth and orthodontic appliances. \hl{All images were acquired by radiologists or dentists with more than five years of experience and labeled by a team of 30 dentists following strict Standard Operating Procedures (SOPs). To ensure consistency, a pilot annotation phase was conducted to unify boundary definitions across the team. A two-stage annotation protocol was employed: each scan was initially annotated by one of ten junior dentists and subsequently reviewed and corrected by one of three senior dentists with over 10 years of experience. To validate the reliability of the ground truth, we assessed the inter-observer agreement on a randomly selected subset of the data. Any remaining discrepancies were resolved through panel consensus.} The dataset has been approved by the Medical Ethics Committee and is licensed under the CC BY-NC-ND license for scientific use only, and commercialization and secondary distribution are prohibited. The data for each task consisted of a small labeled training set and a large unlabeled set used to support the development and evaluation of semi-supervised learning models.

\subsubsection{2D-XRay Dataset}
For Task 1 (2D Panoramic X-ray Dataset), images were categorized into adult and child subsets based on patient age and tooth morphology. During preprocessing, the original DICOM images were converted to PNG format. Dental experts performed instance-level annotations using tools such as EISeg and LabelMe, with a standard image resolution of 640×320 pixels. The annotations, including FDI numbering for each tooth instance, were saved as JSON files. The dataset was divided into training, validation, and test sets, maintaining a realistic clinical ratio of adult to pediatric patients. The training set for Task 1 consisted of 2380 OPGs (30 fully annotated for instance-level segmentation, 2350 unlabeled), with a validation set of 20 OPGs and a test set of 50 OPGs.

\subsubsection{3D-CBCT Dataset}
For Task 2 (3D-CBCT dataset), each 3D volume was meticulously annotated layer by layer by dental experts using ITK-SNAP software. All identifiable teeth (primarily permanent teeth in this task) within each volume were segmented at the instance level and assigned corresponding FDI tooth numbers. The annotation results were saved as nii.gz files. The training set for Task 2 includes 330 CBCT scans (30 of which are fully annotated at the instance level, and 300 are unannotated), the validation set includes 20 CBCT scans, and the test set includes 50 CBCT scans. Among the unlabelled samples, 62.79
\% had artefacts, 54.47\% had fillings, 88.15\% had missing teeth, 92.72\% had implants, 94.59\% had decayed teeth/roots, 97.92\% had deciduous teeth, and 98.34\% had dental braces. Even in the unlabelled training set samples, 35.00\% still contained artefacts. 

\subsection{Participants and challenge phases}
The STS2024 Challenge is aiming to promote the research and practical application of semi-supervised learning methods in medical image analysis.
The challenge is conducted in phases: registration for the competition opens on May 10, 2024, training data is released, and the online validation phase is launched on July 15, test data opens on September 23, and participants are required to submit their final algorithms based on the Docker package before September 26th. The final evaluation results are officially released through the STS2024 Challenge Workshop during the MICCAI 2024 conference, which was held October 6-10, 2024, in Marrakech, Morocco.

The Challenge attracted research teams from several countries and regions around the world, with the largest number of participants from China. According to the participation statistics, Task 1 (2D tooth segmentation task) was participated in by 116 teams with a cumulative number of 346 submissions; \hl{Task 2 (3D CBCT tooth segmentation task) was participated in by 104 teams with a cumulative number of 158 submissions during the preliminary phase. However, to uphold the highest standards of reproducibility, the final leaderboard focuses on the 5 teams that successfully submitted valid Docker containers and open-sourced their code for the hidden test set evaluation.} These data reflect that researchers around the world have paid great attention to tooth image segmentation technology and extensively explored semi-supervised learning methods in real applications. To recognize outstanding contributions, cash prizes of \hl{\$500} were awarded to the top-performing team in each track, with additional souvenirs for other top teams presenting in person.

These data reflect the significant global attention on tooth image segmentation technology and the extensive exploration of semi-supervised learning methods. To foster collaboration and reproducibility, the solutions from the top 10 teams in the 2D track and top 5 in the 3D track are publicly available on our GitHub repository, accompanied by detailed result comparisons. Furthermore, the first authors of the leading teams were invited to co-author this summary paper. Throughout the challenge, participants submitted Dockerized algorithms for evaluation, with a multi-phase validation process allowing them to test and refine their methods multiple times before the final submission, ensuring a fair and robust assessment.

\subsection{Clinical Utility of Segmentation}
The clinical relevance of this challenge arises from its focus on accurate instance-level segmentation of teeth, which has far-reaching clinical implications for various dental specialties. In addition to general identification of tooth regions, the precise localization of each tooth using FDI numbers allows for finer and more accurate diagnosis and treatment.

In orthodontics, precise segmentation of tooth instances~\cite{10820520} is an important prerequisite for diagnosing malocclusion, planning aligner placement paths, and evaluating treatment outcomes. By accurately modeling tooth morphology and spatial relationships, the surgeon can more effectively formulate movement paths and mechanical strategies to enhance the controllability and effectiveness of treatment.

In terms of dental implant surgical planning~\cite{spector2008implant}, instance segmentation can clearly distinguish the target teeth, neighboring tooth roots, and key anatomical structures within the jawbone (e.g., the inferior alveolar nerve), which in turn aids in assessing the bone volume and risk in the implant area. This plays an important role in improving the accuracy of implant position and reducing intraoperative complications.
Additionally, in restorative dentistry and prosthetics, obtaining the precise boundaries of each tooth is the basis for digitally customizing the design of crowns, bridges, and removable prostheses, helping to achieve better occlusal fit and aesthetics.

More broadly, instance-level segmentation~\cite{10.1007/978-3-031-88977-6_15} with AI can also help clinicians visualize lesion progression (e.g., caries, apical periodontitis, etc.), quantitatively assess the evolution of dental disease over time, and even perform 3D visualization simulations prior to complex maxillofacial surgeries. Together, these capabilities drive dental imaging towards intelligent diagnosis and personalized treatment.

\subsection{Performance Evaluation}
The competition requires participants to submit segmentation masks generated on the original test images (in json file format for 2D tasks and .nii.gz for 3D tasks), which are evaluated using a variety of performance and efficiency metrics, including image-level versus instance-level Dice Similarity Coefficients (DSCs), Normalized Surface Distance (NSDs), \hl{and Recognition Accuracy (IAs)}. Algorithm runtime (no more than 60 seconds per case) and GPU memory consumption (based on memory-time curve area) will also be examined.

The pixel-level Dice coefficient is a set similarity measure function that is used to evaluate the degree of similarity between two sets, and its formula is defined as follows:
\begin{equation}
    \text { Dice }_{\text {image }}=\frac{2 *|\mathrm{~A} \cap \mathrm{~B}|}{|\mathrm{A}|+|\mathrm{B}|}
\end{equation}
where A denotes the mask of the proposed model prediction and B denotes the mask of Ground Truth (GT) labeling.


\begin{hlbox}
The pixel-level Normalized Surface Dice (NSD) measures the overall segmentation quality by quantifying the degree of surface overlap between the predicted and ground-truth boundaries. It is defined as follows:

\begin{equation}
    \text{NSD}_{\text{image}} = 
    \frac{\text{overlap}_{\text{GT}} + \text{overlap}_{\text{pred}}}
         {\text{total\_area}_{\text{GT}} + \text{total\_area}_{\text{pred}}},
\end{equation}
\end{hlbox}

\begin{hlbox}
In this equation, $\text{overlap}_{\text{GT}}$ denotes the surface area of the ground-truth boundary that lies within the specified tolerance distance from the predicted boundary, and $\text{overlap}_{\text{pred}}$ represents the surface area of the predicted boundary that lies within the same tolerance distance from the ground-truth boundary. $\text{total\_area}_{\text{GT}}$ and $\text{total\_area}_{\text{pred}}$ refer to the total surface areas (in pixels) of the ground-truth and predicted boundaries, respectively. In this study, the tolerance distance was fixed at 2\,mm, which defines the acceptable spatial deviation between the predicted and ground-truth surfaces. Points on the two surfaces are considered overlapping if their Euclidean distance is less than or equal to 2\,mm. A higher NSD value indicates a greater degree of surface consistency and thus better segmentation accuracy at the boundary level.
\end{hlbox}

\vspace{1em}

\begin{hlbox}
The instance-level Normalized Surface Dice (NSD) evaluates the segmentation performance across multiple individual objects or instances and is defined as follows:

\begin{equation}
\begin{aligned}
    \text{NSD}_{\text{instance}} &= \frac{1}{N} \sum_{i=1}^{N} \text{NSD}_{i}, \\
    \text{where} \quad 
    \text{NSD}_{i} &= 
    \frac{\text{overlap}_{\text{GT}_{i}} + \text{overlap}_{\text{pred}_{i}}}
         {\text{total\_area}_{\text{GT}_{i}} + \text{total\_area}_{\text{pred}_{i}}},
\end{aligned}
\end{equation}

Here, \(N\) denotes the total number of instances in the dataset. For each instance \(i\), $\text{overlap}_{\text{GT}_{i}}$ and $\text{overlap}_{\text{pred}_{i}}$ represent the overlapping areas between the predicted and ground-truth boundaries of that specific instance within the 2\,mm tolerance range, while $\text{total\_area}_{\text{GT}_{i}}$ and $\text{total\_area}_{\text{pred}_{i}}$ denote their respective total surface areas. The instance-level NSD reflects the average boundary accuracy across all segmented instances, ensuring that both global and object-wise geometric consistency are comprehensively evaluated.
\end{hlbox}

To specifically evaluate the critical task of instance identification and separation, we introduced the Instance Affinity (IA) metric. This metric is defined as the fraction of ground truth tooth instances that are correctly detected, where a detection is considered correct if the Intersection over Union (IoU) between the predicted instance and the ground truth instance is greater than or equal to 0.5.
\begin{equation}
    \text{IA} = \frac{\sum_{i=1}^{N_{GT}} \mathbb{I}(\text{max}_{j}(\text{IoU}(G_i, P_j)) \geq 0.5)}{N_{GT}}
\end{equation}
where $G_i$ is the $i$-th ground truth instance, $P_j$ is a predicted instance, $N_{GT}$ is the total number of ground truth instances, and $\mathbb{I}(\cdot)$ is the indicator function. This metric directly penalizes both missed teeth and spurious detections.

\begin{hlbox}
The tooth-level F1 score evaluates the model's ability to correctly detect and segment individual teeth as distinct anatomical instances. It is computed based on connected-component matching between prediction and ground truth, with small components (volume $<$ 11 voxels) discarded as noise. A ground truth tooth is considered correctly detected if at least 65\% of its volume is covered by one or more predicted teeth, each of which contains no more than 70\% non-overlapping (background) voxels. Conversely, a predicted tooth is deemed valid if it satisfies the symmetric matching condition with respect to the ground truth. The tooth-level F1 score is defined as the harmonic mean of tooth-level recall and precision:
\end{hlbox}

\begin{equation}
F_1^{\text{tooth}} =
\begin{cases}
\displaystyle \frac{2 \cdot \text{Precision}_{\text{tooth}} \cdot \text{Recall}_{\text{tooth}}}{\text{Precision}_{\text{tooth}} + \text{Recall}_{\text{tooth}}}, & \text{if } \text{Precision}_{\text{tooth}} + \text{Recall}_{\text{tooth}} > 0, \\
0, & \text{otherwise},
\end{cases}
\end{equation}

\begin{hlbox}
where $\text{Recall}_{\text{tooth}} = \frac{\text{TP}_G}{|G|}$ is the ratio of correctly detected ground truth teeth ($\text{TP}_G$) to the total number of valid ground truth teeth ($|G|$), and $\text{Precision}_{\text{tooth}} = \frac{\text{TP}_P}{|P|}$ is the ratio of valid predicted teeth ($\text{TP}_P$) to the total number of valid predicted teeth ($|P|$). This metric emphasizes instance-wise detection accuracy rather than voxel-wise overlap, making it more clinically relevant for tasks such as individual tooth identification and counting.
\end{hlbox}

In addition to the above metrics, \hl{The competition also considers the algorithm’s runtime and GPU memory consumption, ensuring that the runtime does not exceed 60 seconds in each case and that memory consumption is measured by the area under the memory–time curve. The evaluation process is thoroughly examined using these multi-dimensional metrics to ensure that the algorithms are reasonably evaluated in terms of both performance and efficiency.}

The Challenge attracted research teams from several countries and regions around the world, with the largest number of participants from China. According to the participation statistics, Task 1 (2D tooth segmentation task) was participated in by 116 teams with a cumulative number of 346 submissions; Task 2 (3D CBCT tooth segmentation task) was participated in by 104 teams with a cumulative number of 158 submissions. These data reflect that researchers around the world have paid great attention to tooth image segmentation technology and extensively explored semi-supervised learning methods in real applications. To recognize outstanding contributions, cash prizes of \hl{\$500} were awarded to the top-performing team in each track, with additional souvenirs for other top teams presenting in person.

\section{Methods}
\begin{figure*}[!ht]
\centering
\includegraphics[width=\linewidth]{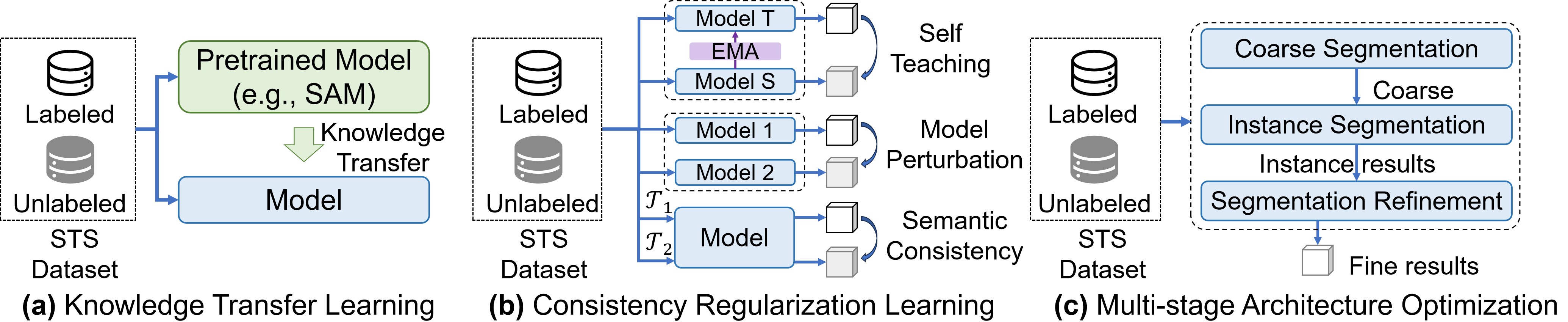}
\caption{Overview of prominent methodological strategies employed by participants in the STS 2024 Challenge. The figure illustrates four key approaches: (a) Knowledge transfer with pretrained models, where pre-trained foundation models (e.g., SAM) are leveraged to improve segmentation. (b) Consistency regularization learning, including self-teaching, model perturbation, and semantic consistency ($\mathcal{T}_{1}$ and $\mathcal{T}_{2}$ denote two kinds of transformation). (c) Multi-stage architecture optimization decomposes the problem into multiple sub-problems and gradually obtained fine results.}
\label{fig:method1}
\end{figure*}

\begin{table*}[t!]
    \centering
    \caption{Summary of 2D submitted teams, model architecture, optimization strategy, and training settings.}
    \label{tab:team_summary}
    \renewcommand{\arraystretch}{1.3} 
    \setlength{\tabcolsep}{4pt}
    \begin{adjustbox}{width=\textwidth}
    \begin{tabular}{lccccccc}
        \toprule
        \textbf{Team} & \makecell[c]{\textbf{Model}\\\textbf{Architecture}} & \textbf{Backbone} & \textbf{Optimizer} & \makecell[c]{\textbf{Loss}\\\textbf{Function}} & \textbf{Device} & \textbf{Epochs} & \makecell[c]{\textbf{Batch}\\\textbf{Size}} \\
        \midrule
        ChohoTech & YOLOv8 & \makecell[c]{YOLOv8} & Adam & \makecell[c]{CIoU, DFL,\\VFL} & RTX 3090 & 100 & 32 \\ \midrule
        Camerart2024 & \makecell[c]{Self-Training\\Pipeline} & DeepLabV3+ & Adam & BCE, Dice & RTX 4090 & 200 & 4 \\ \midrule
        Jichangkai & \makecell[c]{Two-Stage\\Semi-Supervised\\nnU-Net} & nnU-Net & AdamW & Dice, CE & RTX 4090 & 150 & 4 \\ \midrule
        Dew123 & DICL Network & UNet & SGD & \makecell[c]{Dice, MSE,\\CE} & RTX 3060 & 100 & 8 \\ \midrule
        Junqiangmler & Semi-TeethSeg2024 & VNet2d & AdamW & \makecell[c]{Dice,\\Cross-Entropy} & RTX 4090 & 300 & 4 \\  \midrule 
        Isjinghao & SemiT-SAM & SAM & AdamW & \makecell[c]{Multi-\\component} & RTX 3060 & 300 & 4 \\ \midrule 
        Lazyman & \makecell[c]{Cross Teaching\\Network} & \makecell[c]{CNN +\\Transformer} & SGD & Dice, MSE & RTX 4090 & 43 & 16 \\ \midrule 
        Caiyichen & YOLOv9 & \makecell[c]{YOLOv9} & Adam & \makecell[c]{CIoU, DFL,\\VFL} & RTX 3060 & 100 & 16 \\ \midrule 
        Guo7777 & \makecell[c]{ResUnet50\\+ SAM} & \makecell[c]{ResNet50,\\SAM} & Adam & \makecell[c]{BCEWithLogitsLoss,\\MSELoss} & \makecell[c]{Tesla V100\\-SXM2} & 300 & 4 \\ \midrule 
        Ccc2024 & DAE-Net & \makecell[c]{Dual Attention\\Mechanism} & Adam & Dice, IoU & RTX 4060 Ti & 40 & 32 \\
        \bottomrule
    \end{tabular}
    \end{adjustbox}
\end{table*}

\begin{table*}[t!]
    \centering
    \caption{Summary of 3D submitted teams, model architecture, optimization strategy, and training settings for STS MICCAI 2024 Challenge Task 2.}
    \label{tab:team_summary_3d_combined_final}
    \renewcommand{\arraystretch}{1.3} 
    \setlength{\tabcolsep}{4pt} 
    \begin{adjustbox}{width=\textwidth}
    \begin{tabular}{llccccc} 
        \toprule
        \textbf{Team} & \makecell[c]{\textbf{Model}\\\textbf{Architecture}} & \textbf{Backbone} & \textbf{Optimizer} & \makecell[c]{\textbf{Loss}\\\textbf{Function}} & \textbf{Device} & \makecell[c]{\textbf{Epochs}}\ \\
        \midrule
        Chohotech & \makecell[c]{3-Stage\\Pipeline} & \makecell[c]{YOLOv8,\\U-Net} & Adam & \makecell[c]{CIoU, DFL,\\VFL} & NVIDIA A100 & 100 \\
        \midrule
        Houwentai & \makecell[c]{CFP 2-Stage\\Semi-sup.\\nnU-Net} & \makecell[c]{nnU-Net} & AdamW & Dice, CE & 6 $\times$ RTX 4090 & 100 \\
        \midrule
        Madongdong & \makecell[c]{Semi-supervised\\YOLOv8} & YOLOv8 & Adam & \makecell[c]{CIoU, DFL,\\VFL} & RTX 3090 & 300 \\
        \midrule
        Jichangkai & \makecell[c]{2-Stage nnU-Net\\Self-training} & nnU-Net & AdamW & Dice, CE & RTX 3090 & 300 \\
        \midrule
        Junqiangmler & \makecell[c]{ROI Preproc.\\+ VNet3d} & VNet3d & AdamW  & Dice, CE & RTX 4090 & 300 \\
        \midrule
        Gute\_iici & \makecell[c]{2-Stage\\Unimatch} & \makecell[c]{VNet (S1),\\Enc-Dec (S2)} & AdamW & Unimatch & Tesla V100 & 100  \\
        \bottomrule
    \end{tabular}
    \end{adjustbox}
\end{table*}

\begin{figure}[!ht]
    \centering
    \includegraphics[width=\linewidth]{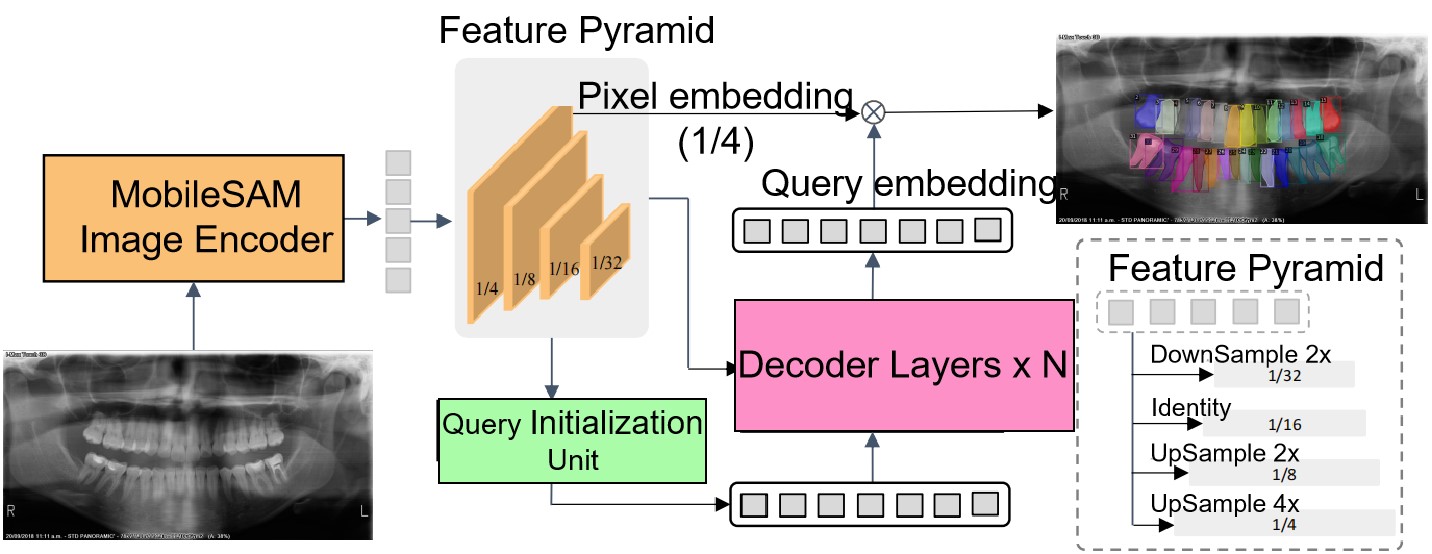}
    \caption{Architecture of the SemiT-SAM model, submitted by team 'Isjinhao' for the 2D challenge track. The model employs an encoder-decoder structure comprising an image encoder, a basic feature pyramid, a query initialization unit, and a mask decoder.}
    \label{fig:2d_rank6}
\end{figure}

\begin{hlbox}
\subsection{SAM-based Knowledge Transfer}
In the dental segmentation task, major challenges include the scarcity of labeled data and the anatomical complexity of teeth. To address these issues, participants integrated the Segment Anything Model (SAM) to leverage its powerful general segmentation capabilities and edge sensitivity.

\textbf{Chohotech} combined YOLOv8~\cite{redmon2016you} and SAM~\cite{kirillov2023segment} in a semi-supervised pipeline. YOLOv8 first generates bounding boxes to prompt SAM for fine pixel-level segmentation, creating high-quality pseudo-labels. These labels are refined via a quality filtering mechanism and used for iterative optimization. The method also adapts feature resolutions and loss functions for dental images and employs the Hungarian algorithm to resolve tooth number matching.

\textbf{Guo777} proposed a framework integrating ResNet50~\cite{he2016deep} and SAM-Med2D~\cite{cheng2023sam}. The ResNet50 encoder and U-Net decoder extract image features, while the SAM-Med2D module utilizes an attention mechanism to focus on key anatomical regions. \hl{To adapt to the intensity differences in medical X-rays, the input images were preprocessed using min-max normalization to rescale pixel values to [0, 1] and resized to a consistent 512×512 resolution. This standardization ensures robust feature extraction despite varying lighting and exposure conditions.} A semi-supervised strategy further expands the training set through pseudo-label generation and screening.

\textbf{Isjinhao} introduced SemiT-SAM, an innovative visual base model. Inspired by SAM-style architectures, it inherits MobileSAM’s lightweight ViT-Tiny backbone and produces multi-scale feature maps (C2–C5) through a simple feature pyramid. \hl{To handle the spatial variability of dental radiographs, images were resized while preserving the aspect ratio (max dimension 1024) and zero-padded to a fixed 1024×1024 input size, avoiding geometric distortion.} In the semi-supervised phase, a teacher-student distillation strategy is used: the teacher generates pseudo-labels, which are \hl{filtered using class-confidence and mask-size thresholds to ensure high-quality supervision, following the SAM-based distillation workflow.} The student model is jointly trained on labeled and pseudo-labeled data, optimizing the teacher weights via Exponential Moving Averaging (EMA) (Fig. \ref{fig:2d_rank6}).
\end{hlbox}

\begin{figure}[!ht]
    \centering
    \includegraphics[width=\linewidth]{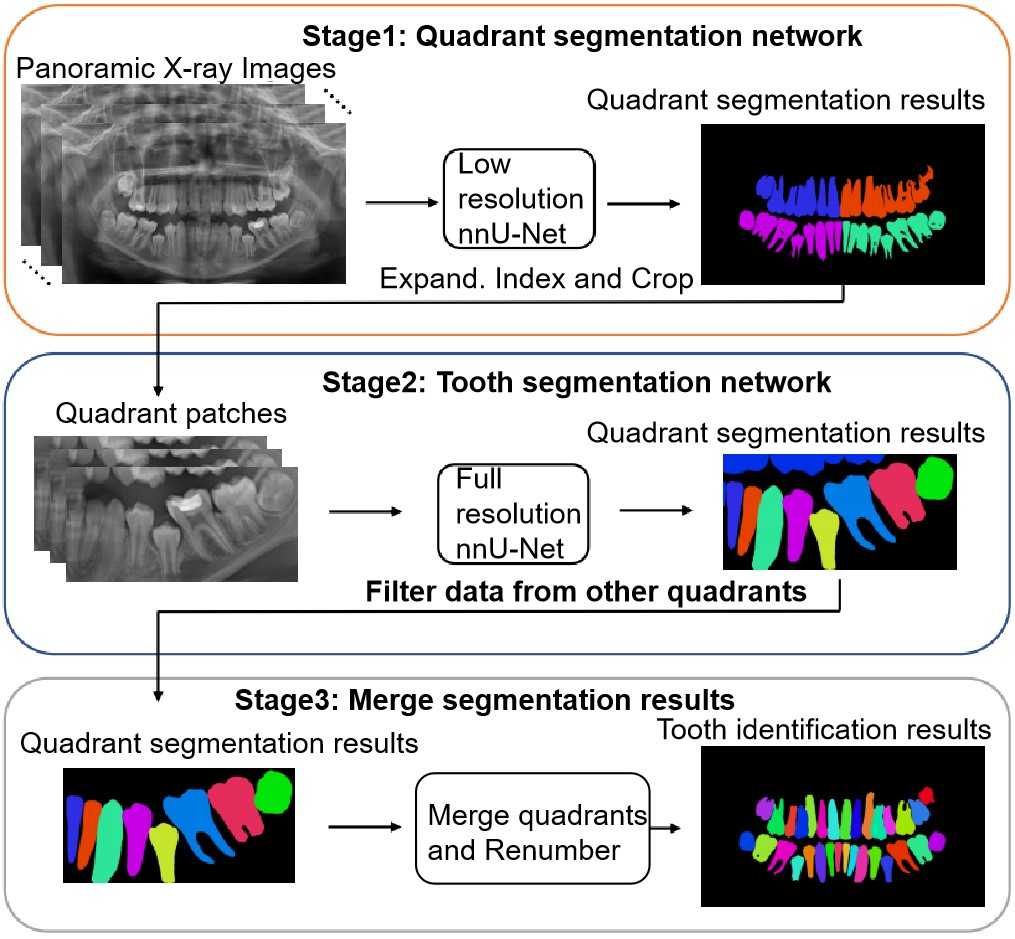}
    \caption{Schematic diagram of the two-stage segmentation method submitted by team 'Jichangkai' for the 2D challenge track. The first stage divides all teeth into four quadrants. The second stage identifies and segments each tooth in the quadrants. The third stage combines the results from all four quadrants to reconstruct the segmentation in the original image space.}
    \label{fig:2d_rank3}
\end{figure}

\begin{hlbox}
\subsection{Semantic Consistency Learning}
Consistency learning enhances model robustness by ensuring feature and semantic alignment between labeled and unlabeled data, often through perturbations and multi-scale learning.

\textbf{Houwentai} proposed a Coarse-to-Fine Pseudo-labeling (CFP) method. An initial model trained on limited labeled data generates coarse pseudo-labels, from which high-confidence samples are filtered for fine segmentation training. The architecture improves upon the 3D nnU-Net~\cite{isensee2021nnu} by incorporating residual modules for stability. Test-time augmentation (TTA) is applied during inference to integrate multiple predictions, significantly boosting accuracy and robustness under limited supervision.

\textbf{Jichangkai} developed a two-stage semi-supervised framework for both 2D and 3D tasks. In Stage 1, images are segmented into four anatomical quadrants using a low-resolution nnU-Net to simplify the spatial context. In Stage 2, each quadrant is processed by a full-resolution nnU-Net for detailed instance segmentation. A teacher-student framework iteratively generates and filters pseudo-labels, while specific mechanisms remove interfering data from adjacent quadrants. Finally, results are merged and renumbered to produce the complete segmentation (Fig. \ref{fig:2d_rank3}).
\end{hlbox}

\begin{figure}[!ht]
    \centering
    \includegraphics[width=\linewidth]{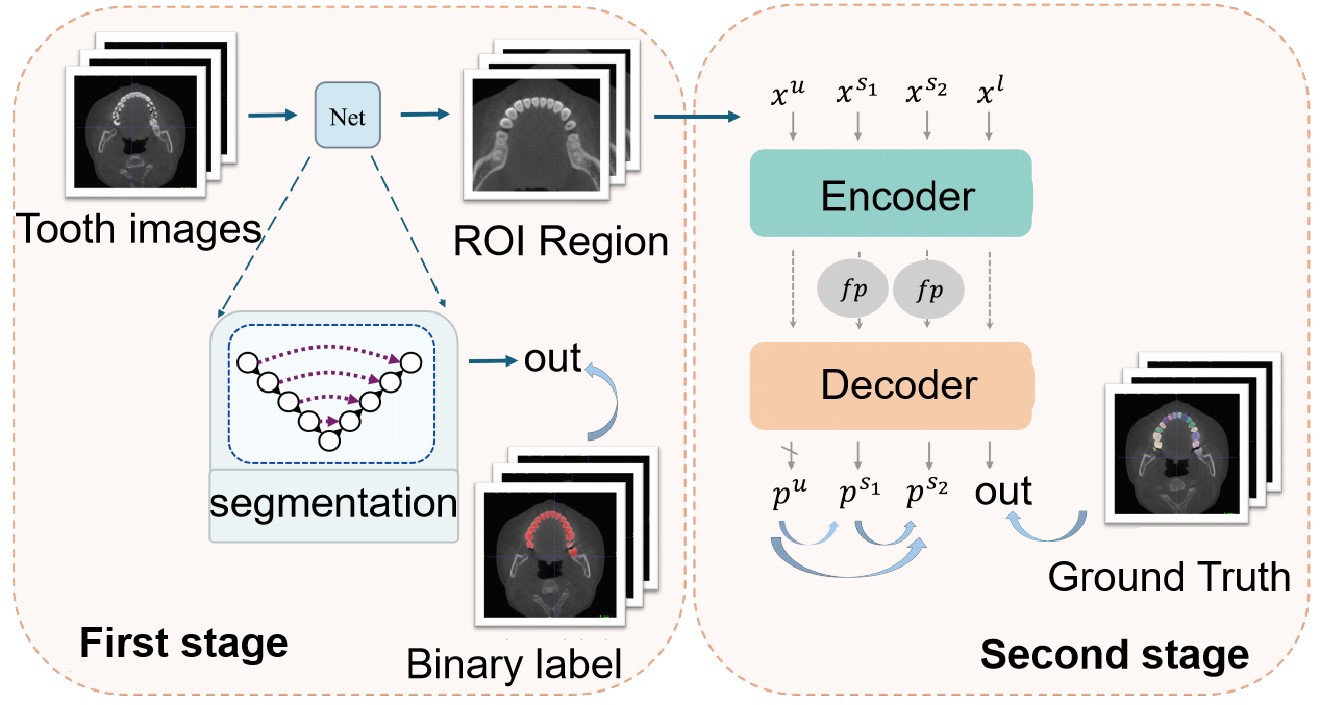}
    \caption{Two-stage semi-supervised tooth segmentation framework proposed by team 'Gute-iici' for the 3D challenge track. Unimatch is utilized for semi-supervised learning in the second stage.}
    \label{fig:3d_rank6} 
\end{figure}

\begin{hlbox}
\textbf{Camerart} utilized a self-training approach for 2D segmentation. High-quality pseudo-labels are generated via multi-model integration and morphological manipulation, refined through 5-fold cross-validation. The DeepLabV3+ model employs a sigmoid function to handle tooth overlaps. During inference, segmentation is optimized using dual inference (original and flipped images) and contour extraction.

\textbf{Lazyman} combined U-Net~\cite{ronneberger2015u} and Swin-UNet~\cite{cao2022swin} in a cross-network co-training mechanism. U-Net captures local edge details, while Swin-UNet leverages Transformers for global context. The networks generate pseudo-labels for each other, enabling the model to learn from unlabeled data by enforcing consistency between local and global feature representations.

\textbf{Dew123} proposed a deformable intrinsic consistency learning method. A semi-supervised stage uses a deformable convolutional module and cross-attention to generate pseudo-labels, followed by a fully-supervised stage combining these with labeled data. Semantic consistency is enforced via a composite loss function (Dice, cross-entropy, and consistency loss).

\textbf{Gute-iici} applied the Unimatch framework in a two-stage 3D pipeline. A V-Net~\cite{milletari2016v} first performs binary segmentation to extract ROIs. Subsequently, a multi-head network performs binary and multi-category segmentation. Feature perturbation at the bottleneck layer allows unenhanced data to supervise strongly enhanced outputs, improving feature learning from unlabeled data (Fig. \ref{fig:3d_rank6}).
\end{hlbox}

\begin{figure}[!ht]
    \centering
    \includegraphics[width=\linewidth]{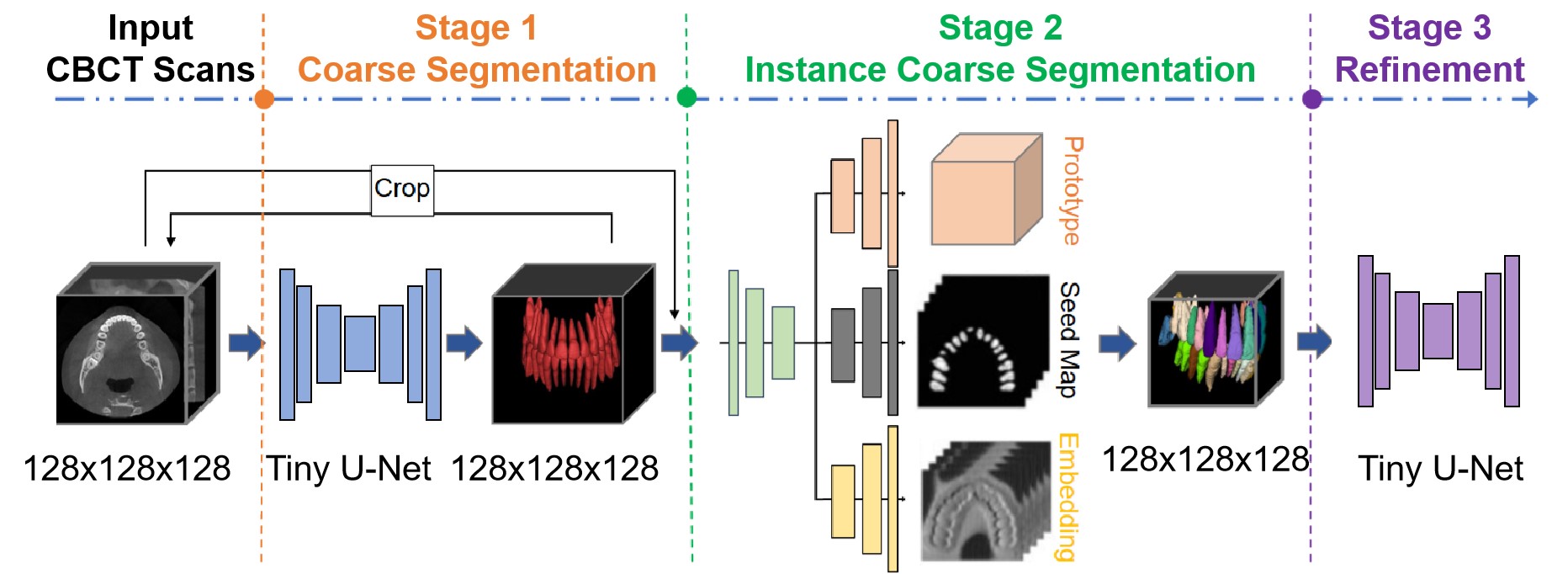}
    \caption{Overview of the T3 Net method, submitted by team 'Haoyuuuu' for the 3D challenge track. This automated pipeline performs instance-level segmentation and numbering of teeth and implants in CBCT images using a cascaded three-stage network.}
    \label{fig:3d_rank5}
\end{figure}

\begin{hlbox}
\subsection{Coarse-to-Fine Optimization}
To handle the complexity of dental structures, several teams adopted multi-stage strategies that progressively refine segmentation from coarse localization to fine-grained detailing.

\textbf{Haoyuuu} proposed T3Net, a three-stage framework for CBCT images. Stage 1 uses a simplified Tiny V-Net~\cite{milletari2016v} for coarse semantic segmentation to localize the ROI. Stage 2 employs a modified 3D ERFNet with spatial embedding, seed map, and prototype learning branches to generate coarse instance masks. Stage 3 refines these masks using a full-resolution Tiny V-Net. This cascaded approach effectively resolves under- and over-segmentation issues in complex 3D data (Fig. \ref{fig:3d_rank5}).
\end{hlbox}

\begin{hlbox}
\textbf{Madongdong} optimized the YOLOv8 architecture~\cite{redmon2016you} by introducing a coarse-to-fine structure, a decoupled head, and an anchor-free design. Their multi-stage training involves initially training on labeled data to generate pseudo-labels, which are then used to extend the dataset for a second stage of fine-tuning. This iterative process allows the model parameters to be optimized for high precision even with scarce labeled data.

\textbf{Chohotech (3D)} adapted their 2D strategy to CBCT data. The method begins with 2D Maximum Intensity Projections (MIP) processed by YOLOv8 for fast ROI detection. A modified 3D YOLOv8 then segments tooth instances within the ROI, using an adapted anchor strategy. Finally, a U-Net refines tooth boundaries using a combination of binary cross-entropy and Dice losses. This synergistic approach significantly improves efficiency and accuracy in volumetric segmentation.
\end{hlbox}

\begin{table*}[t!]
    \centering
    \caption{
    Quantitative results for the 2D Panoramic X-ray (PXI) track. 
    The top table shows the overall performance across multiple metrics including std in parentheses. 
    The bottom table details the Instance DSC scores broken down by anatomical quadrant and patient age group (adult vs. child). 
    The elevated standard deviations observed in certain teams may be attributed to the presence of a few cases with extreme complexity, 
    as indicated in Fig.~\ref{fig:rainplot_complete} and Fig.~\ref{fig:relation_2d_01}. 
    AUC\_GPU : Total GPU memory usage integrated over time, reflecting overall resource consumption. 
    \textbf{F1}: The instance-level F1 score evaluates the balance between precision and recall for correctly detected individual tooth instances.
    }
    \label{table:results_2d}
    \renewcommand{\arraystretch}{1.4} 
    \begin{adjustbox}{width=\textwidth}
        \begin{tabular}{lc|cc|cccc|cc}
        \toprule
        \multicolumn{2}{c|}{\textbf{Team}} & \multicolumn{2}{c|}{\textbf{Image-level}} & \multicolumn{4}{c|}{\textbf{Instance-level}} & \multicolumn{2}{c}{\textbf{Algorithm-level}}\\
        \midrule
        Name & Ranking 
        & \makecell[c]{DSC $\uparrow$} 
        & \makecell[c]{NSD $\uparrow$} 
        & \makecell[c]{DSC $\uparrow$} 
        & \makecell[c]{NSD $\uparrow$} 
        & \makecell[c]{IA $\uparrow$} 
        & \colorbox{yellow}{\makecell[c]{F1 $\uparrow$}} 
        & \makecell[c]{AUC\_GPU \\ (GB$\cdot$s) $\downarrow$} 
        & \makecell[c]{RT \\ (s) $\downarrow$} \\
        \midrule
        ChohoTech & 1 & \makecell[c]{92.03 \\ (\scriptsize{$\pm$}2.36)} & \makecell[c]{95.89 \\ (\scriptsize{$\pm$}1.74)} & \makecell[c]{83.59 \\ (\scriptsize{$\pm$}15.01)} & \makecell[c]{87.52 \\ (\scriptsize{$\pm$}15.47)} & \makecell[c]{88.53 \\ (\scriptsize{$\pm$}18.81)} & \colorbox{yellow}{\makecell[c]{92.54 \\ (\scriptsize{$\pm$}14.07)}} & \makecell[c]{7341.12 \\ (\scriptsize{$\pm$}1449.20)} & \makecell[c]{13.29 \\ (\scriptsize{$\pm$}2.93)} \\ \midrule
        Camerart2024 & 2 & \makecell[c]{87.20 \\ (\scriptsize{$\pm$}4.58)} & \makecell[c]{91.01 \\ (\scriptsize{$\pm$}4.70)} & \makecell[c]{70.75 \\ (\scriptsize{$\pm$}16.87)} & \makecell[c]{74.42 \\ (\scriptsize{$\pm$}17.56)} & \makecell[c]{70.01 \\ (\scriptsize{$\pm$}21.71)} & \colorbox{yellow}{\makecell[c]{80.07 \\ (\scriptsize{$\pm$}18.44)}} & \makecell[c]{14250.98 \\ (\scriptsize{$\pm$}3870.70)} & \makecell[c]{13.27 \\ (\scriptsize{$\pm$}3.67)} \\
        \midrule
        Jichangkai & 3 & \makecell[c]{92.01 \\ (\scriptsize{$\pm$}13.24)} & \makecell[c]{94.87 \\ (\scriptsize{$\pm$}13.63)} & \makecell[c]{78.98 \\ (\scriptsize{$\pm$}21.13)} & \makecell[c]{81.93 \\ (\scriptsize{$\pm$}21.74)} & \makecell[c]{80.06 \\ (\scriptsize{$\pm$}24.91)} & \colorbox{yellow}{\makecell[c]{85.98 \\ (\scriptsize{$\pm$}21.98)}} & \makecell[c]{25461.26 \\ (\scriptsize{$\pm$}1159.91)} & \makecell[c]{55.90 \\ (\scriptsize{$\pm$}2.15)} \\
        \midrule
        Dew123 & 4 & \makecell[c]{87.94 \\ (\scriptsize{$\pm$}4.69)} & \makecell[c]{92.09 \\ (\scriptsize{$\pm$}4.53)} & \makecell[c]{69.24 \\ (\scriptsize{$\pm$}21.58)} & \makecell[c]{73.49 \\ (\scriptsize{$\pm$}22.42)} & \makecell[c]{65.93 \\ (\scriptsize{$\pm$}25.71)} & \colorbox{yellow}{\makecell[c]{75.88 \\ (\scriptsize{$\pm$}23.94)}} & \makecell[c]{15088.50 \\ (\scriptsize{$\pm$}4448.06)} & \makecell[c]{13.91 \\ (\scriptsize{$\pm$}4.49)} \\
        \midrule
        Junqiangmler & 5 & \makecell[c]{82.72 \\ (\scriptsize{$\pm$}10.13)} & \makecell[c]{86.85 \\ (\scriptsize{$\pm$}10.53)} & \makecell[c]{64.02 \\ (\scriptsize{$\pm$}18.43)} & \makecell[c]{68.00 \\ (\scriptsize{$\pm$}19.27)} & \makecell[c]{55.20 \\ (\scriptsize{$\pm$}24.15)} & \colorbox{yellow}{\makecell[c]{67.57 \\ (\scriptsize{$\pm$}23.56)}} & \makecell[c]{12483.38 \\ (\scriptsize{$\pm$}3398.61)} & \makecell[c]{14.05 \\ (\scriptsize{$\pm$}3.96)} \\
        \midrule
        Isjinghao & 6 & \makecell[c]{82.64 \\ (\scriptsize{$\pm$}18.37)} & \makecell[c]{86.34 \\ (\scriptsize{$\pm$}18.87)} & \makecell[c]{64.71 \\ (\scriptsize{$\pm$}23.51)} & \makecell[c]{67.75 \\ (\scriptsize{$\pm$}24.28)} & \makecell[c]{67.31 \\ (\scriptsize{$\pm$}26.69)} & \colorbox{yellow}{\makecell[c]{76.68 \\ (\scriptsize{$\pm$}24.48)}} & \makecell[c]{27987.90 \\ (\scriptsize{$\pm$}3297.90)} & \makecell[c]{21.13 \\ (\scriptsize{$\pm$}3.66)} \\
        \midrule
        Lazyman & 7 & \makecell[c]{59.76 \\ (\scriptsize{$\pm$}9.36)} & \makecell[c]{87.05 \\ (\scriptsize{$\pm$}13.61)} & \makecell[c]{49.22 \\ (\scriptsize{$\pm$}13.18)} & \makecell[c]{72.60 \\ (\scriptsize{$\pm$}19.34)} & \makecell[c]{8.57 \\ (\scriptsize{$\pm$}7.82)} & \colorbox{yellow}{\makecell[c]{14.50 \\ (\scriptsize{$\pm$}12.27)}} & \makecell[c]{13910.32 \\ (\scriptsize{$\pm$}5098.61)} & \makecell[c]{11.81 \\ (\scriptsize{$\pm$}3.81)} \\
        \midrule
        Caiyichen & 8 & \makecell[c]{90.80 \\ (\scriptsize{$\pm$}2.53)} & \makecell[c]{94.52 \\ (\scriptsize{$\pm$}2.40)} & \makecell[c]{57.70 \\ (\scriptsize{$\pm$}27.60)} & \makecell[c]{60.48 \\ (\scriptsize{$\pm$}28.82)} & \makecell[c]{51.58 \\ (\scriptsize{$\pm$}33.84)} & \colorbox{yellow}{\makecell[c]{60.59 \\ (\scriptsize{$\pm$}34.32)}} & \makecell[c]{26666.57 \\ (\scriptsize{$\pm$}2129.66)} & \makecell[c]{19.53 \\ (\scriptsize{$\pm$}1.16)} \\
        \midrule
        Guo77777 & 9 & \makecell[c]{75.48 \\ (\scriptsize{$\pm$}9.44)} & \makecell[c]{80.30 \\ (\scriptsize{$\pm$}9.51)} & \makecell[c]{35.84 \\ (\scriptsize{$\pm$}13.04)} & \makecell[c]{38.63 \\ (\scriptsize{$\pm$}13.61)} & \makecell[c]{27.04 \\ (\scriptsize{$\pm$}16.43)} & \colorbox{yellow}{\makecell[c]{39.98 \\ (\scriptsize{$\pm$}20.82)}} & \makecell[c]{19694.26 \\ (\scriptsize{$\pm$}3798.62)} & \makecell[c]{18.42 \\ (\scriptsize{$\pm$}3.18)} \\
        \midrule
        Cccc2024 & 10 & \makecell[c]{91.59 \\ (\scriptsize{$\pm$}2.30)} & \makecell[c]{95.15 \\ (\scriptsize{$\pm$}2.02)} & \makecell[c]{26.60 \\ (\scriptsize{$\pm$}15.20)} & \makecell[c]{27.75 \\ (\scriptsize{$\pm$}15.71)} & \makecell[c]{15.38 \\ (\scriptsize{$\pm$}16.10)} & \colorbox{yellow}{\makecell[c]{24.09 \\ (\scriptsize{$\pm$}19.24)}} & \makecell[c]{17730.48 \\ (\scriptsize{$\pm$}1915.22)} & \makecell[c]{13.46 \\ (\scriptsize{$\pm$}1.48)} \\
        \midrule
        \multicolumn{2}{c|}{baseline} & \makecell[c]{89.38 \\ (\scriptsize{$\pm$}3.81)} & \makecell[c]{51.61 \\ (\scriptsize{$\pm$}1.13)} & \makecell[c]{71.12 \\ (\scriptsize{$\pm$}14.69)} & \makecell[c]{41.45 \\ (\scriptsize{$\pm$}8.56)} & \makecell[c]{44.25 \\ (\scriptsize{$\pm$}34.27)} & \colorbox{yellow}{\makecell[c]{55.54 \\ (\scriptsize{$\pm$}37.18)}} & \makecell[c]{11104.75  \\ (\scriptsize{$\pm$}4882.34)} & \makecell[c]{1.22 \\ (\scriptsize{$\pm$}0.09)} \\
        \bottomrule
        \end{tabular}
    \end{adjustbox}
\end{table*}



\begin{table*}[t!]
    \centering
    \caption{Quantitative results for the 2D Panoramic X-ray (PXI) track at quadrant level. The details include the Instance DSC scores broken down by anatomical quadrant and patient age group (adult vs. child).}
    \label{table:results_2d_quadrant}
    \renewcommand{\arraystretch}{1.4} 
    \begin{adjustbox}{width=\textwidth}
        \begin{tabular}{lc|cc|cc|cc|cc|c}
        \toprule
        \multicolumn{2}{c|}{\textbf{Quadrant}} & \multicolumn{2}{c|}{\makecell[c]{\textbf{Upper left}}} & \multicolumn{2}{c|}{\makecell[c]{\textbf{Upper right}}} & \multicolumn{2}{c|}{\makecell[c]{\textbf{Lower right}}} & \multicolumn{2}{c}{\makecell[c]{\textbf{Lower left}}} & \multicolumn{1}{|c}{\makecell[c]{\textbf{Total}}} \\
        \midrule
        Team Name & Ranking & Adult & Child & Adult & Child & Adult & Child & Adult & Child & All age groups  \\
        \midrule
        ChohoTech & 1 & \makecell[c]{83.45 \\ (\scriptsize{$\pm$}22.19)} & \makecell[c]{84.87 \\ (\scriptsize{$\pm$}19.58)} & \makecell[c]{85.09 \\ (\scriptsize{$\pm$}21.27)} & \makecell[c]{81.77 \\ (\scriptsize{$\pm$}25.44)} & \makecell[c]{78.24 \\ (\scriptsize{$\pm$}5.81)} & \makecell[c]{76.25 \\ (\scriptsize{$\pm$}12.23)} & \makecell[c]{80.42 \\ (\scriptsize{$\pm$}7.99)} & \makecell[c]{80.35 \\ (\scriptsize{$\pm$}6.34)} & \makecell[c]{82.68 \\ (\scriptsize{$\pm$}20.14)} \\
        \midrule
        Jichangkai & 3 & \makecell[c]{78.09 \\ (\scriptsize{$\pm$}28.71)} & \makecell[c]{79.09 \\ (\scriptsize{$\pm$}26.14)} & \makecell[c]{80.32 \\ (\scriptsize{$\pm$}30.61)} & \makecell[c]{78.32 \\ (\scriptsize{$\pm$}30.64)} & \makecell[c]{76.84 \\ (\scriptsize{$\pm$}10.57)} & \makecell[c]{74.72 \\ (\scriptsize{$\pm$}13.27)} & \makecell[c]{79.37 \\ (\scriptsize{$\pm$}11.18)} & \makecell[c]{77.30 \\ (\scriptsize{$\pm$}11.44)} & \makecell[c]{78.53 \\ (\scriptsize{$\pm$}26.23)} \\
        \midrule
        Dew123 & 4 & \makecell[c]{69.93 \\ (\scriptsize{$\pm$}21.30)} & \makecell[c]{70.77 \\ (\scriptsize{$\pm$}25.55)} & \makecell[c]{75.57 \\ (\scriptsize{$\pm$}24.21)} & \makecell[c]{73.27 \\ (\scriptsize{$\pm$}26.93)} & \makecell[c]{57.46 \\ (\scriptsize{$\pm$}15.00)} & \makecell[c]{57.06 \\ (\scriptsize{$\pm$}16.47)} & \makecell[c]{60.29 \\ (\scriptsize{$\pm$}7.42)} & \makecell[c]{61.67 \\ (\scriptsize{$\pm$}10.04)} & \makecell[c]{69.41 \\ (\scriptsize{$\pm$}23.25)} \\
        \midrule
        Camerart2024 & 2 & \makecell[c]{69.88 \\ (\scriptsize{$\pm$}25.13)} & \makecell[c]{71.06 \\ (\scriptsize{$\pm$}22.95)} & \makecell[c]{69.30 \\ (\scriptsize{$\pm$}26.30)} & \makecell[c]{68.63 \\ (\scriptsize{$\pm$}24.94)} & \makecell[c]{59.95 \\ (\scriptsize{$\pm$}11.73)} & \makecell[c]{69.78 \\ (\scriptsize{$\pm$}13.80)} & \makecell[c]{71.81 \\ (\scriptsize{$\pm$}6.92)} & \makecell[c]{66.34 \\ (\scriptsize{$\pm$}13.28)} & \makecell[c]{69.10 \\ (\scriptsize{$\pm$}22.73)} \\
        \midrule
        Junqiangmler & 5 & \makecell[c]{62.07 \\ (\scriptsize{$\pm$}21.95)} & \makecell[c]{61.66 \\ (\scriptsize{$\pm$}23.25)} & \makecell[c]{67.55 \\ (\scriptsize{$\pm$}24.86)} & \makecell[c]{70.03 \\ (\scriptsize{$\pm$}24.63)} & \makecell[c]{61.67 \\ (\scriptsize{$\pm$}17.35)} & \makecell[c]{58.53 \\ (\scriptsize{$\pm$}14.39)} & \makecell[c]{68.53 \\ (\scriptsize{$\pm$}9.34)} & \makecell[c]{63.28 \\ (\scriptsize{$\pm$}13.46)} & \makecell[c]{64.81 \\ (\scriptsize{$\pm$}22.20)} \\
        \midrule
        Isjinghao & 6 & \makecell[c]{64.59 \\ (\scriptsize{$\pm$}26.93)} & \makecell[c]{74.24 \\ (\scriptsize{$\pm$}28.02)} & \makecell[c]{66.79 \\ (\scriptsize{$\pm$}31.65)} & \makecell[c]{67.30 \\ (\scriptsize{$\pm$}27.54)} & \makecell[c]{46.23 \\ (\scriptsize{$\pm$}20.96)} & \makecell[c]{52.98 \\ (\scriptsize{$\pm$}16.03)} & \makecell[c]{49.95 \\ (\scriptsize{$\pm$}16.62)} & \makecell[c]{50.46 \\ (\scriptsize{$\pm$}17.10)} & \makecell[c]{64.12 \\ (\scriptsize{$\pm$}27.84)} \\
        \midrule
        Caiyichen & 8 & \makecell[c]{60.33 \\ (\scriptsize{$\pm$}30.55)} & \makecell[c]{54.46 \\ (\scriptsize{$\pm$}34.00)} & \makecell[c]{47.40 \\ (\scriptsize{$\pm$}35.58)} & \makecell[c]{60.37 \\ (\scriptsize{$\pm$}30.37)} & \makecell[c]{49.72 \\ (\scriptsize{$\pm$}23.86)} & \makecell[c]{44.18 \\ (\scriptsize{$\pm$}32.28)} & \makecell[c]{41.36 \\ (\scriptsize{$\pm$}28.38)} & \makecell[c]{48.31 \\ (\scriptsize{$\pm$}24.99)} & \makecell[c]{53.45 \\ (\scriptsize{$\pm$}32.27)} \\
        \midrule
        Lazyman & 8 & \makecell[c]{52.65 \\ (\scriptsize{$\pm$}14.88)} & \makecell[c]{52.20 \\ (\scriptsize{$\pm$}17.58)} & \makecell[c]{47.83 \\ (\scriptsize{$\pm$}17.60)} & \makecell[c]{50.64 \\ (\scriptsize{$\pm$}17.98)} & \makecell[c]{39.13 \\ (\scriptsize{$\pm$}14.27)} & \makecell[c]{40.40 \\ (\scriptsize{$\pm$}10.01)} & \makecell[c]{46.63 \\ (\scriptsize{$\pm$}7.08)} & \makecell[c]{44.04 \\ (\scriptsize{$\pm$}8.44)} & \makecell[c]{48.97 \\ (\scriptsize{$\pm$}16.31)} \\
        \midrule
        Guo77777 & 9 & \makecell[c]{48.59 \\ (\scriptsize{$\pm$}24.99)} & \makecell[c]{52.79 \\ (\scriptsize{$\pm$}21.17)} & \makecell[c]{43.49 \\ (\scriptsize{$\pm$}22.69)} & \makecell[c]{0.99 \\ (\scriptsize{$\pm$}3.03)} & \makecell[c]{43.04 \\ (\scriptsize{$\pm$}17.74)} & \makecell[c]{29.66 \\ (\scriptsize{$\pm$}17.98)} & \makecell[c]{27.04 \\ (\scriptsize{$\pm$}12.91)} & \makecell[c]{13.86 \\ (\scriptsize{$\pm$}9.39)} & \makecell[c]{34.66 \\ (\scriptsize{$\pm$}27.00)} \\
        \midrule
        Cccc2024 & 10 & \makecell[c]{38.63 \\ (\scriptsize{$\pm$}21.17)} & \makecell[c]{15.10 \\ (\scriptsize{$\pm$}25.27)} & \makecell[c]{6.49 \\ (\scriptsize{$\pm$}8.79)} & \makecell[c]{40.15 \\ (\scriptsize{$\pm$}19.68)} & \makecell[c]{37.39 \\ (\scriptsize{$\pm$}14.32)} & \makecell[c]{18.97 \\ (\scriptsize{$\pm$}13.73)} & \makecell[c]{18.62 \\ (\scriptsize{$\pm$}17.19)} & \makecell[c]{42.66 \\ (\scriptsize{$\pm$}13.93)} & \makecell[c]{26.06 \\ (\scriptsize{$\pm$}23.36)} \\
        \midrule
        \multicolumn{2}{c|}{baseline} & \makecell[c]{48.56 \\ (\scriptsize{$\pm$}28.64)} & \makecell[c]{36.37 \\ (\scriptsize{$\pm$}23.09)} & \makecell[c]{50.36 \\ (\scriptsize{$\pm$}32.27)} & \makecell[c]{38.99 \\ (\scriptsize{$\pm$}26.25)} & \makecell[c]{50.05 \\ (\scriptsize{$\pm$}34.35)} & \makecell[c]{48.08 \\ (\scriptsize{$\pm$}27.04)} & \makecell[c]{52.40 \\ (\scriptsize{$\pm$}34.94)} & \makecell[c]{48.35 \\ (\scriptsize{$\pm$}25.35)} & \makecell[c]{48.68 \\ (\scriptsize{$\pm$}31.46)} \\
        \bottomrule
        \end{tabular}
    \end{adjustbox}
\end{table*}

\begin{figure*}[!ht]
    \centering
    \includegraphics[width=\linewidth]{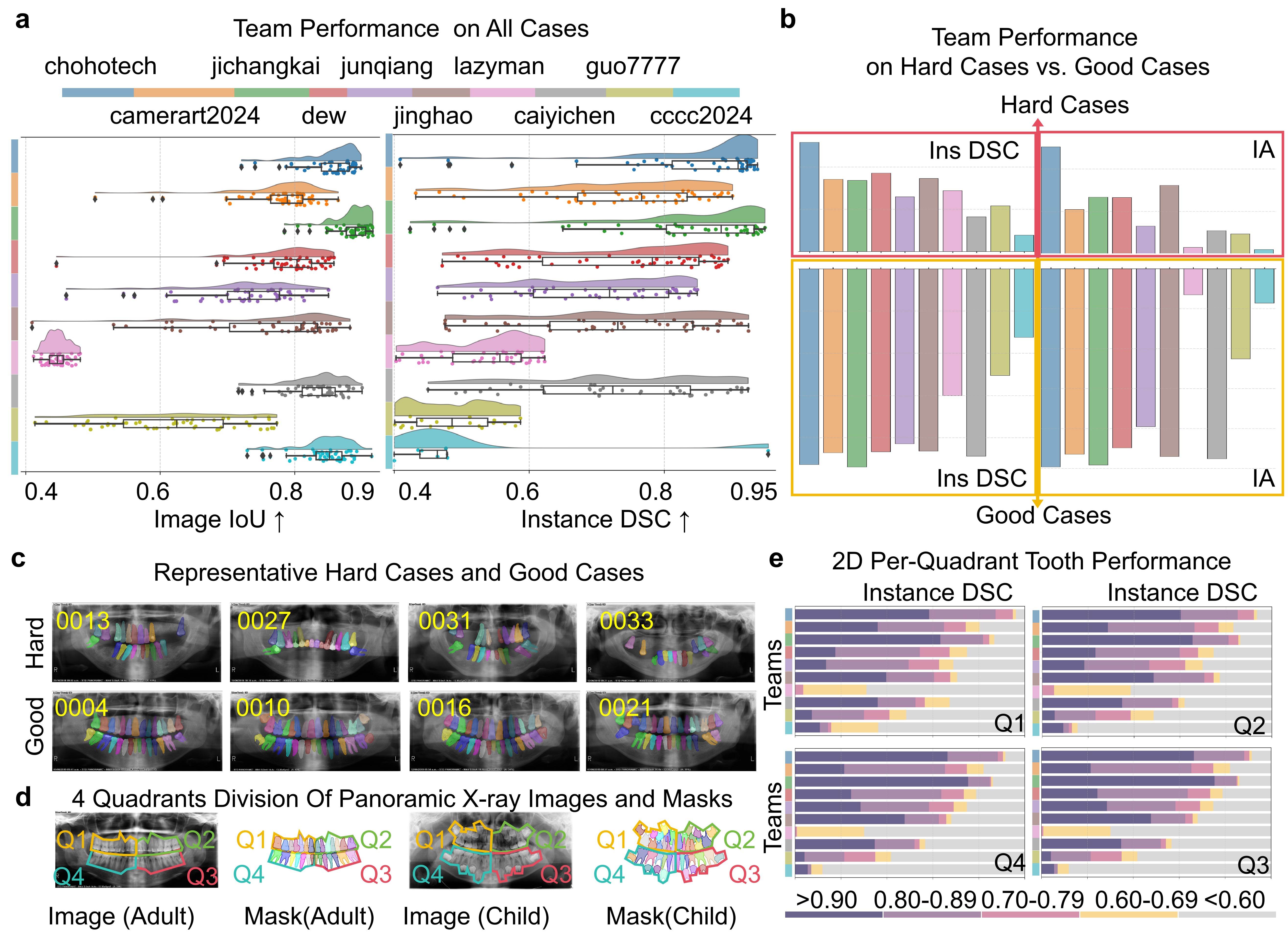}
    \caption{Analysis of 2D challenge track results. (a) Statistical summary of segmentation metrics for the top 10 teams on the test set. (b) Team performance on challenging cases and well-segmented cases. (c) Comparison of challenging cases with poor segmentation performance across most methods versus well-segmented cases, highlighting differences in tooth integrity. (d) Illustration of the four-quadrant division of a panoramic X-ray. (e) Instance DSC score distributions across four quadrants for participating teams. Quadrants: Q1 (Upper Right), Q2 (Upper Left), Q3 (Lower Left), Q4 (Lower Right). Ins DSC denotes Instance DSC.}    
    \label{fig:rainplot_complete}
\end{figure*}

\subsection{Quantitative Evaluation}
\subsubsection{Quantitative Evaluation in 2D PXI Track}

The comprehensive performance of the top ten teams in the 2D Panoramic X-ray (PXI) track, detailed in Table \ref{table:results_2d}, underscores the advantage of semi-supervised learning (SSL) for this complex instance segmentation task. The most crucial finding of this challenge is the substantial performance gain achieved by SSL methods over a conventional supervised baseline \hl{within the constrained data regime of the challenge. We acknowledge that supervised learning remains the gold standard when large-scale annotations are available. However, in this specific scenario where the training set was restricted to just 30 labeled images (approx. 1\% of the total dataset), SSL provided a critical performance boost.}\textbf{} When trained only on the 30 labeled images, a fully-supervised nnU-Net baseline achieved an Instance Affinity (IA) of 44.17\%. In stark contrast, the top-performing SSL method from team ChohoTech reached an IA of 88.53\%, representing a relative improvement of over 100\% (or an absolute gain of 44.36 percentage points). This result powerfully validates the central hypothesis of our challenge: that leveraging large amounts of unlabeled data through SSL is essential for achieving high performance in instance-level dental segmentation.

The leading methods employed diverse yet effective SSL strategies. The top-ranked team, ChohoTech, utilized a YOLOv8-based detection-then-segmentation pipeline, achieving a robust instance-level Dice of 83.59\%. Other leading approaches included a self-training framework (Camerart2024, rank 2) and a two-stage method built upon the powerful nnU-Net baseline (Jichangkai, rank 3). The raincloud plot in Fig. \ref{fig:rainplot_complete}(a) visually confirms a clear performance stratification, with the top-tier teams consistently outperforming the lower-ranked participants in instance-level metrics, highlighting the efficacy of these advanced SSL frameworks.

The detailed quadrant-level analysis in Table \ref{table:results_2d_quadrant} and the visualization in Fig. \ref{fig:rainplot_complete}(e) show that most teams achieved relatively stable performance across the four anatomical quadrants (Q1: Upper Right, Q2: Upper Left, Q3: Lower Left, Q4: Lower Right). This consistency is likely attributable to the inherent anatomical symmetry of human dentition, which aids model generalization.
\begin{hlbox}
    The analysis of challenging versus well-segmented cases in Fig. \ref{fig:rainplot_complete}(b-c) further clarifies performance variations across "normal" and "abnormal" cohorts. As shown in Fig. \ref{fig:rainplot_complete} (b) and (c), well-segmented cases, such as 0004, 0010, 0016, and 0021, typically featured complete and well-aligned dentition. In contrast, challenging cases (e.g., 0013, 0027, 0031, 0033) were often characterized by severe tooth loss (edentulism) or complex pathologies. The performance on these "abnormal" cases degraded significantly for nearly all teams, as shown in Fig. \ref{fig:rainplot_complete} (b) by the lower and more variable scores.
\end{hlbox}

Finally, the results highlight the nuanced trade-offs between different SSL strategies. For instance, the performance of the team Jichangkai reveals the importance of a multi-metric evaluation. Despite achieving high image-level and instance-level Dice scores, their comparatively lower IA score (80.06\% vs. the winner's 88.53\%) suggests that while their nnU-Net based method produced spatially accurate masks, it struggled more with separating heavily overlapping teeth compared to the detection-first approach of ChohoTech. This underscores that different SSL architectures may excel at different aspects of the instance segmentation task—spatial accuracy versus topological correctness—providing valuable insights for future methods development.

\begin{figure*}[!ht]
    \centering
    \includegraphics[width=\linewidth]{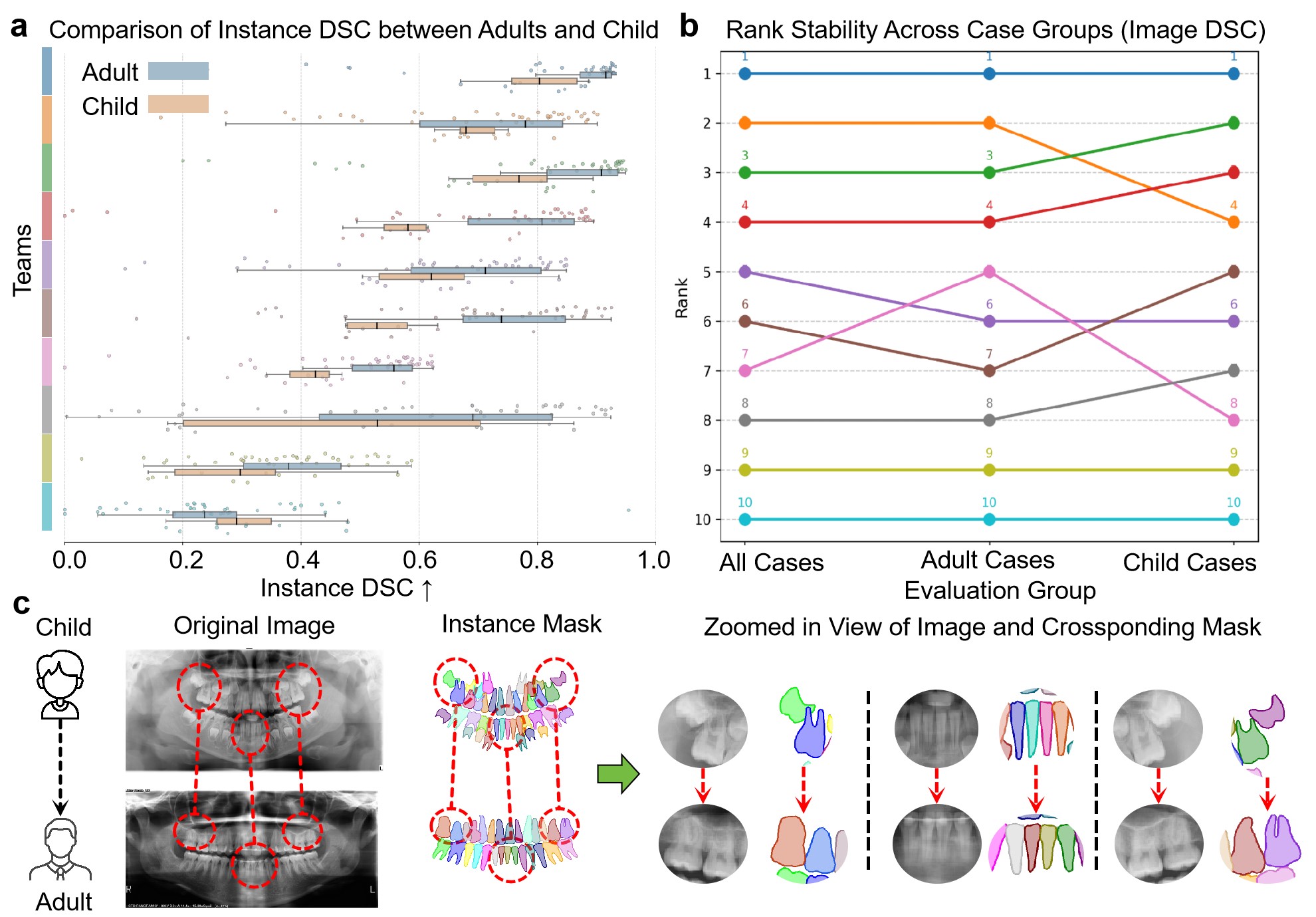}
    \caption{Performance analysis based on age groups in the 2D challenge track. (a) Statistical comparison of segmentation metrics for the top 10 teams on adult versus pediatric cases. (b) Illustration of anatomical differences and similarities between adult and pediatric dentition in Orthopantomogram. (c) Comparison of segmentation stability (e.g., variance in performance) across adult and pediatric cases for participating teams.}
    \label{fig:rainplot_adultandcild}
\end{figure*}

\subsubsection{Performance on Challenging Pediatric and Adult Cohorts}
To assess the robustness of the submitted SSL methods, we analyzed their performance when stratified by patient age (adult vs. child), a key challenge highlighted in this work. The results reveal that while pediatric cases are inherently more difficult, leading SSL methods demonstrated remarkable generalization capabilities.

As visualized in the raincloud plot in Fig. \ref{fig:rainplot_adultandcild}(a-b), a discernible performance gap exists between the two cohorts, with mean Instance DSC scores for adult cases generally being higher and less variable than for pediatric cases. This gap is attributable to both the smaller number of labeled pediatric examples in the training set and the intrinsic complexity of pediatric dentition. As shown in Fig. \ref{fig:rainplot_adultandcild}(c), pediatric anatomy often features smaller, more densely packed teeth with obvious overlap between deciduous and developing permanent teeth, making the instance segmentation task fundamentally more challenging.
\begin{hlbox}
The rank stability analysis in Fig. \ref{fig:rainplot_adultandcild}(b) further underscores this challenge. To quantify the robustness of the rankings against demographic shifts, we calculated Spearman's Rank Correlation Coefficient between team rankings on adult and pediatric cohorts.
The analysis yielded a correlation coefficient of $\rho=0.879$ ($p=0.000814$) between adult and pediatric rankings, indicating strong overall consistency. While this demonstrates robustness of the top-performing methods across different age groups, several teams' rankings did fluctuate significantly between the two cohorts, highlighting the challenge-specific performance variations. The correlations between overall rankings and age-specific rankings were $\rho=0.964$ ($p=7.32\times10^{-6}$) for the adult cohort and $\rho=0.939$ ($p=5.48\times10^{-5}$) for the pediatric cohort, confirming the stability of our evaluation framework. This indicates that their SSL strategies were not equally effective at generalizing across these distinct demographics. However, the quadrant-level results in Table \ref{table:results_2d_quadrant} show that top-performing teams like ChohoTech and Jichangkai maintained impressive and relatively stable performance across all quadrants for both age groups. 
\end{hlbox} 
This indicates that their SSL strategies were not equally effective at generalizing across these distinct demographics. However, the quadrant-level results in Table \ref{table:results_2d_quadrant} show that top-performing teams like ChohoTech and Jichangkai maintained impressive and relatively stable performance across all quadrants for both age groups.

Crucially, the success of the top methods on this difficult, underrepresented pediatric data is a significant outcome of this challenge. It validates the central premise that semi-supervised learning is a highly effective paradigm for developing robust models in clinical scenarios where acquiring labeled data for every subgroup is impractical. The ability to generalize from a few labeled examples to complex, unlabeled pediatric cases demonstrates a critical step towards building clinically viable and equitable AI tools.

\begin{table*}[t!]
    \centering
    \caption{Quantitative results for the 3D Dental CBCT track. The top table shows the overall performance across multiple metrics. The bottom table details the Instance DSC and NSD scores for each anatomical quadrant. The increased standard deviations in certain teams may stem from the inclusion of a limited number of cases with exceptional complexity, as evidenced by the results in Fig. \ref{fig:relation_2d_01}.}
    \label{table:results_3d}
    \renewcommand{\arraystretch}{1.4} 
    \begin{adjustbox}{width=\textwidth}
        \begin{tabular}{lc|cc|cccc|cc}
        \toprule
        \multicolumn{2}{c|}{\textbf{Team}} 
        & \multicolumn{2}{c|}{\textbf{Image-level}} 
        & \multicolumn{4}{c|}{\textbf{Instance-level}} 
        & \multicolumn{2}{c}{\textbf{Algorithm-level}}\\
        \midrule
        Name & Ranking 
        & \makecell[c]{DSC $\uparrow$} 
        & \makecell[c]{NSD $\uparrow$} 
        & \makecell[c]{DSC $\uparrow$} 
        & \makecell[c]{NSD $\uparrow$} 
        & \makecell[c]{IA $\uparrow$} 
        & \colorbox{yellow}{\makecell[c]{F1 $\uparrow$}} 
        & \makecell[c]{AUC\_GPU \\ (GB$\cdot$s) $\downarrow$} 
        & \makecell[c]{RT \\ (s) $\downarrow$}  \\
        \midrule
        ChohoTech & 1 
        & \makecell[c]{93.54 \\ (\scriptsize{$\pm$}1.48)} 
        & \makecell[c]{97.42 \\ (\scriptsize{$\pm$}1.35)} 
        & \makecell[c]{92.15 \\ (\scriptsize{$\pm$}2.26)} 
        & \makecell[c]{96.60 \\ (\scriptsize{$\pm$}2.34)} 
        & \makecell[c]{98.39 \\ (\scriptsize{$\pm$}2.32)} 
        & \colorbox{yellow}{\makecell[c]{99.56 \\ (\scriptsize{$\pm$}1.03)}}
        & \makecell[c]{233660.20 \\ (\scriptsize{$\pm$}119552.16)} 
        & \makecell[c]{60.76 \\ (\scriptsize{$\pm$}19.22)} \\
        \midrule
        Houwentai & 2 
        & \makecell[c]{76.79 \\ (\scriptsize{$\pm$}32.13)} 
        & \makecell[c]{78.52 \\ (\scriptsize{$\pm$}32.08)} 
        & \makecell[c]{83.59 \\ (\scriptsize{$\pm$}33.82)} 
        & \makecell[c]{73.77 \\ (\scriptsize{$\pm$}34.51)} 
        & \makecell[c]{75.24 \\ (\scriptsize{$\pm$}36.59)} 
        & \colorbox{yellow}{\makecell[c]{99.38 \\ (\scriptsize{$\pm$}2.11)}}
        & \makecell[c]{829283.02 \\ (\scriptsize{$\pm$}308410.35)} 
        & \makecell[c]{210.37 \\ (\scriptsize{$\pm$}53.56)} \\
        \midrule
        Madongdong & 3 
        & \makecell[c]{83.57 \\ (\scriptsize{$\pm$}2.78)} 
        & \makecell[c]{80.87 \\ (\scriptsize{$\pm$}5.82)} 
        & \makecell[c]{77.87 \\ (\scriptsize{$\pm$}5.36)} 
        & \makecell[c]{73.63 \\ (\scriptsize{$\pm$}7.58)} 
        & \makecell[c]{88.18 \\ (\scriptsize{$\pm$}10.64)} 
        & \colorbox{yellow}{\makecell[c]{93.63 \\ (\scriptsize{$\pm$}6.78)}}
        & \makecell[c]{48266.68 \\ (\scriptsize{$\pm$}19234.55)} 
        & \makecell[c]{52.82 \\ (\scriptsize{$\pm$}17.99)} \\
        \midrule
        Jichangkai & 4 
        & \makecell[c]{76.50 \\ (\scriptsize{$\pm$}29.64)} 
        & \makecell[c]{85.34 \\ (\scriptsize{$\pm$}19.77)} 
        & \makecell[c]{73.29 \\ (\scriptsize{$\pm$}32.22)} 
        & \makecell[c]{78.81 \\ (\scriptsize{$\pm$}28.20)} 
        & \makecell[c]{72.44 \\ (\scriptsize{$\pm$}39.12)} 
        & \colorbox{yellow}{\makecell[c]{76.05 \\ (\scriptsize{$\pm$}37.21)}}
        & \makecell[c]{377331.12 \\ (\scriptsize{$\pm$}260326.42)} 
        & \makecell[c]{214.72 \\ (\scriptsize{$\pm$}110.78)} \\
        \midrule
        Junqiangmler & 5 
        & \makecell[c]{77.40 \\ (\scriptsize{$\pm$}26.22)} 
        & \makecell[c]{77.39 \\ (\scriptsize{$\pm$}27.61)} 
        & \makecell[c]{65.83 \\ (\scriptsize{$\pm$}31.04)} 
        & \makecell[c]{66.74 \\ (\scriptsize{$\pm$}31.67)} 
        & \makecell[c]{65.56 \\ (\scriptsize{$\pm$}36.29)} 
        & \colorbox{yellow}{\makecell[c]{72.07 \\ (\scriptsize{$\pm$}36.98)}}
        & \makecell[c]{1004507.86 \\ (\scriptsize{$\pm$}528550.02)} 
        & \makecell[c]{114.11 \\ (\scriptsize{$\pm$}51.96)} \\
        \midrule
        \multicolumn{2}{c|}{baseline} 
        & \makecell[c]{71.99 \\ (\scriptsize{$\pm$}35.90)} 
        & \makecell[c]{73.34 \\ (\scriptsize{$\pm$}37.12)} 
        & \makecell[c]{30.80 \\ (\scriptsize{$\pm$}22.29)} 
        & \makecell[c]{37.28 \\ (\scriptsize{$\pm$}23.72)} 
        & \makecell[c]{14.98 \\ (\scriptsize{$\pm$}23.29)} 
        & \colorbox{yellow}{\makecell[c]{18.04 \\ (\scriptsize{$\pm$}22.91)}}
        & \makecell[c]{5089073.15 \\ (\scriptsize{$\pm$}3453290.65)} 
        & \makecell[c]{130.21 \\ (\scriptsize{$\pm$}62.19)} \\
        \bottomrule
        \end{tabular}
    \end{adjustbox}
\end{table*}


\begin{table*}[t!]
    \centering
    \caption{Quantitative results for the 3D Dental CBCT track at quadrant level. The details include the Instance DSC and NSD scores for each anatomical quadrant.}
    \label{table:results_3d_quadrant}
    \renewcommand{\arraystretch}{1.4} 
    \begin{adjustbox}{width=\textwidth}
        \begin{tabular}{lc|cc|cc|cc|cc|c}
        \toprule
        \multicolumn{2}{c|}{\textbf{Quadrant}} & \multicolumn{2}{c|}{\makecell[c]{\textbf{Upper left}}} & \multicolumn{2}{c|}{\makecell[c]{\textbf{Upper right}}} & \multicolumn{2}{c|}{\makecell[c]{\textbf{Lower right}}} & \multicolumn{2}{c}{\makecell[c]{\textbf{Lower left}}} & \multicolumn{1}{|c}{\makecell[c]{\textbf{Total}}} \\
        \midrule
        Team Name & Ranking 
        & \makecell[c]{Dice $\uparrow$} 
        & \makecell[c]{NSD $\uparrow$} 
        & \makecell[c]{Dice $\uparrow$} 
        & \makecell[c]{NSD $\uparrow$} 
        & \makecell[c]{Dice $\uparrow$} 
        & \makecell[c]{NSD $\uparrow$} 
        & \makecell[c]{Dice $\uparrow$} 
        & \makecell[c]{NSD $\uparrow$} 
        & \makecell[c]{Total Dice $\uparrow$} \\
        \midrule
        ChohoTech & 1 & \makecell[c]{93.43 \\ (\scriptsize{$\pm$}1.79)} & \makecell[c]{99.72 \\ (\scriptsize{$\pm$}0.75)} & \makecell[c]{93.56 \\ (\scriptsize{$\pm$}1.76)} & \makecell[c]{99.87 \\ (\scriptsize{$\pm$}0.44)} & \makecell[c]{93.41 \\ (\scriptsize{$\pm$}1.72)} & \makecell[c]{99.92 \\ (\scriptsize{$\pm$}0.14)} & \makecell[c]{93.28 \\ (\scriptsize{$\pm$}1.70)} & \makecell[c]{99.88 \\ (\scriptsize{$\pm$}0.52)} & \makecell[c]{93.56 \\ (\scriptsize{$\pm$}1.49)} \\
        \midrule
        Houwentai & 2 & \makecell[c]{95.78 \\ (\scriptsize{$\pm$}2.84)} & \makecell[c]{99.74 \\ (\scriptsize{$\pm$}1.59)} & \makecell[c]{95.97 \\ (\scriptsize{$\pm$}1.13)} & \makecell[c]{99.98 \\ (\scriptsize{$\pm$}0.05)} & \makecell[c]{95.10 \\ (\scriptsize{$\pm$}2.77)} & \makecell[c]{99.63 \\ (\scriptsize{$\pm$}1.12)} & \makecell[c]{94.67 \\ (\scriptsize{$\pm$}4.11)} & \makecell[c]{99.41 \\ (\scriptsize{$\pm$}2.34)} & \makecell[c]{95.66 \\ (\scriptsize{$\pm$}1.47)} \\
        \midrule
        Madongdong & 3 & \makecell[c]{82.68 \\ (\scriptsize{$\pm$}3.12)} & \makecell[c]{98.38 \\ (\scriptsize{$\pm$}2.55)} & \makecell[c]{84.54 \\ (\scriptsize{$\pm$}2.75)} & \makecell[c]{97.72 \\ (\scriptsize{$\pm$}2.11)} & \makecell[c]{82.64 \\ (\scriptsize{$\pm$}5.62)} & \makecell[c]{95.84 \\ (\scriptsize{$\pm$}5.99)} & \makecell[c]{81.08 \\ (\scriptsize{$\pm$}4.31)} & \makecell[c]{96.09 \\ (\scriptsize{$\pm$}3.37)} & \makecell[c]{84.09 \\ (\scriptsize{$\pm$}2.92)} \\
        \midrule
        Jichangkai & 4 & \makecell[c]{77.08 \\ (\scriptsize{$\pm$}32.09)} & \makecell[c]{85.57 \\ (\scriptsize{$\pm$}27.08)} & \makecell[c]{71.43 \\ (\scriptsize{$\pm$}35.48)} & \makecell[c]{81.78 \\ (\scriptsize{$\pm$}29.13)} & \makecell[c]{72.67 \\ (\scriptsize{$\pm$}35.81)} & \makecell[c]{78.11 \\ (\scriptsize{$\pm$}33.23)} & \makecell[c]{75.12 \\ (\scriptsize{$\pm$}35.88)} & \makecell[c]{81.17 \\ (\scriptsize{$\pm$}32.05)} & \makecell[c]{76.54 \\ (\scriptsize{$\pm$}29.92)}\\
        \midrule
        Junqiangmler & 5 & \makecell[c]{79.94 \\ (\scriptsize{$\pm$}29.40)} & \makecell[c]{86.00 \\ (\scriptsize{$\pm$}30.51)} & \makecell[c]{80.12 \\ (\scriptsize{$\pm$}28.45)} & \makecell[c]{86.22 \\ (\scriptsize{$\pm$}29.52)} & \makecell[c]{69.49 \\ (\scriptsize{$\pm$}26.84)} & \makecell[c]{76.74 \\ (\scriptsize{$\pm$}27.96)} & \makecell[c]{73.55 \\ (\scriptsize{$\pm$}26.42)} & \makecell[c]{81.71 \\ (\scriptsize{$\pm$}27.58)} & \makecell[c]{77.42 \\ (\scriptsize{$\pm$}26.46)} \\
        \midrule
       \multicolumn{2}{c|}{baseline} & \makecell[c]{36.17 \\ (\scriptsize{$\pm$}23.55)} & \makecell[c]{40.76 \\ (\scriptsize{$\pm$}25.13)} & \makecell[c]{32.05 \\ (\scriptsize{$\pm$}24.55)} & \makecell[c]{38.51 \\ (\scriptsize{$\pm$}25.58)} & \makecell[c]{34.85 \\ (\scriptsize{$\pm$}25.35)} &  \makecell[c]{43.18 \\ (\scriptsize{$\pm$}25.70)} & \makecell[c]{33.60 \\ (\scriptsize{$\pm$}26.54)} & \makecell[c]{39.04 \\ (\scriptsize{$\pm$}28.98)} & \makecell[c]{71.98 \\ (\scriptsize{$\pm$}35.88)} \\
        \bottomrule
        \end{tabular}
    \end{adjustbox}
\end{table*}

\begin{figure*}[!ht]
    \centering
    \includegraphics[width=\linewidth]{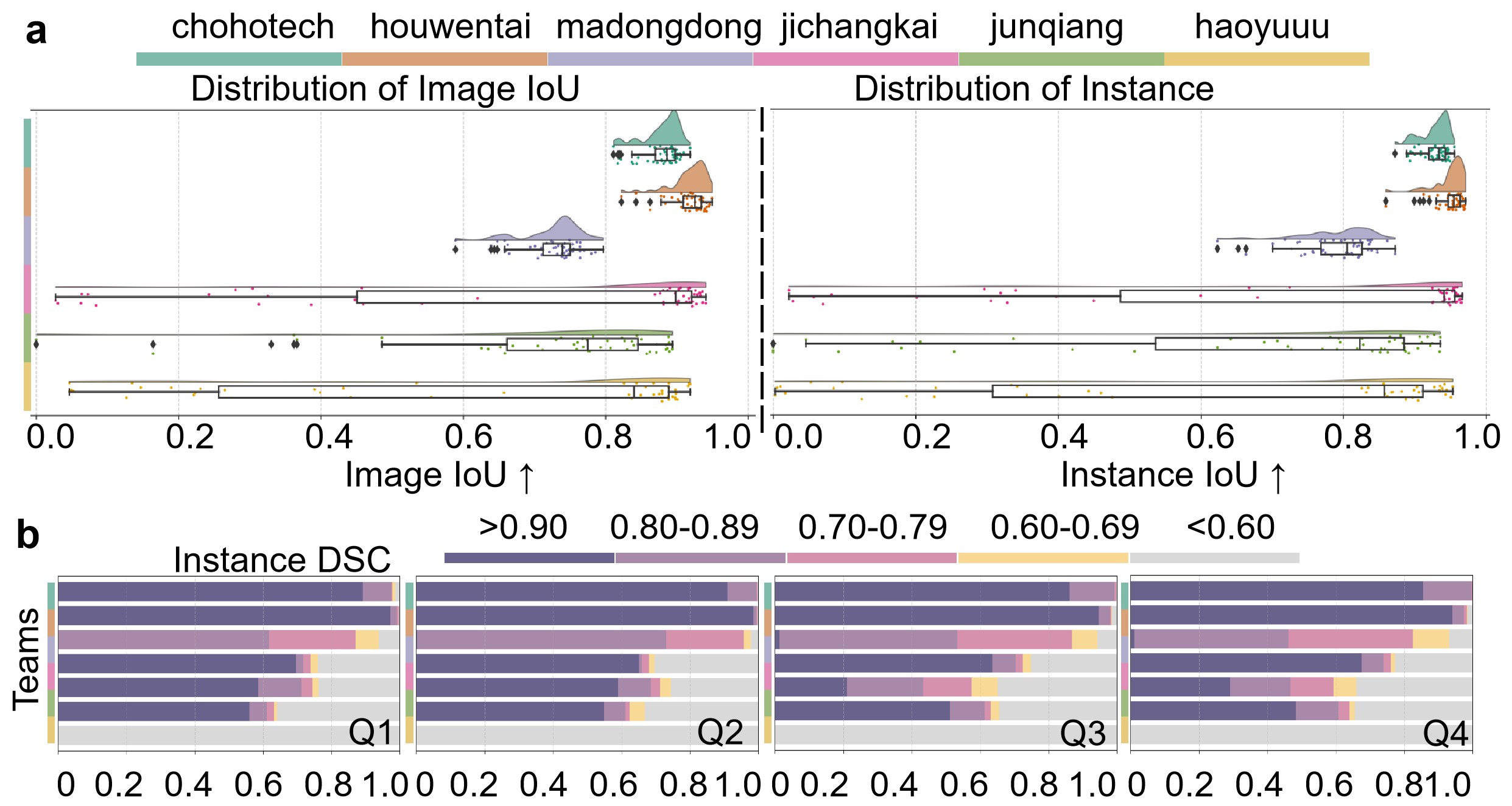}
    \caption{Summary and qualitative assessment of 3D challenge track results. (a) Statistical summary of test set metrics for the top 7 teams. (b) Distribution of Instance DSC scores for multiple teams across the four anatomical quadrants in CBCT.}
    \label{fig:rainplot_3DCBCT}
\end{figure*}

\subsubsection{Quantitative Evaluation in 3D CBCT Track}

The 3D CBCT segmentation track posed a significant challenge, characterized by high-dimensional data and a high prevalence of imaging artifacts. This complexity was reflected in the final submissions, with only five teams successfully completing the task. The results, however, provide a powerful demonstration of semi-supervised learning's efficacy in this demanding volumetric domain.

The most striking outcome of the 3D track is the dramatic performance improvement conferred by SSL. A fully-supervised 3D nnU-Net baseline, trained only on the 30 labeled CBCT scans, achieved a modest Instance DSC of 30.80\%. In contrast, the winning SSL method from team ChohoTech achieved an Instance DSC of 92.15\%. This represents an absolute gain of over 61 percentage points (a relative improvement of 197\%), decisively validating the use of SSL for volumetric dental instance segmentation and proving its ability to overcome extreme data scarcity.

As detailed in Table \ref{table:results_3d} and visualized in Fig. \ref{fig:rainplot_3DCBCT}(a), team ChohoTech's approach, which adapted a YOLO-based pipeline to 3D, was not only the most accurate but also the most stable, evidenced by its low standard deviations across all metrics. This stability contrasts sharply with the performance of several other teams, including the second-place winner Houwentai, which exhibited very high variance (e.g., a standard deviation of 33.82\% for Instance DSC). This suggests that their SSL strategies, while effective on some cases, struggled to generalize across the full diversity of the unlabeled test set. This instability may indicate that their pseudo-labeling or consistency schemes were sensitive to domain shifts introduced by clinical factors like metal artifacts, causing degraded performance on unlabeled data that differed significantly from the small labeled set.

The quadrant-level analysis in Table \ref{table:results_3d_quadrant} and Fig. \ref{fig:rainplot_3DCBCT}(b) further reinforces the robustness of the top methods, which maintained consistently high performance across all four anatomical quadrants. The clinical relevance of these results is underscored by the excellent Normalized Surface Distance (NSD) scores achieved by the top teams. A high NSD score indicates the generation of segmentations with highly accurate surface boundaries, which is a critical prerequisite for clinical applications such as the digital design and fabrication of precise orthodontic appliances and surgical guides.

\subsubsection{Comparative Analysis between 2D and 3D Tracks}

A cross-task comparison reveals differences in performance, model complexity, and computational cost between the 2D PXI and 3D CBCT tracks, highlighting the distinct challenges of each modality. Four teams: ChohoTech, Jichangkai, Junqiangmler, and Madongdong competed in both tracks, allowing for a direct comparison of how their strategies adapted across dimensions.

The most striking difference lies in the potential for segmentation accuracy. As shown in Table \ref{table:results_2d} and Table \ref{table:results_3d}, the winning team ChohoTech achieved a remarkable Instance DSC of 92.15\% in the 3D track, substantially higher than their already impressive 83.59\% in the 2D track. This suggests that the additional spatial information in 3D data is highly effective at resolving ambiguities inherent in 2D projections, most notably the superposition of tooth structures that complicates boundary delineation, leading to more accurate and robust instance segmentation once a model can effectively process volumetric data. However, capitalizing on this 3D advantage proved non-trivial. Other teams like Jichangkai and Junqiangmler saw a decrease in their average scores when moving from 2D to 3D, underscoring the challenge of successfully adapting SSL methods to a higher-dimensional space.

This difficulty is intrinsically linked to the algorithmic efficiency, where the computational burden of the 3D track was an order of magnitude greater. For ChohoTech, the average runtime (RT) increased from 13.29 seconds per 2D image to 60.76 seconds per 3D volume. More dramatically, their integrated GPU memory usage (AUC\_GPU) surged from approximately 7,341 GB$\cdot$s to 233,660 GB$\cdot$s. This massive increase in resource requirements explains why fewer teams were able to complete the 3D task and emphasizes a critical trade-off for clinical deployment: while 3D models can yield superior accuracy, their high computational cost may limit their practical application. The semi-supervised learning paradigm itself is also more demanding in 3D, as generating and refining pseudo-labels for volumetric data requires significantly more memory and computational power, a factor that contributes to the performance instability observed in several 3D submissions.

\begin{figure*}[!ht]
    \centering
    \includegraphics[width=\linewidth]{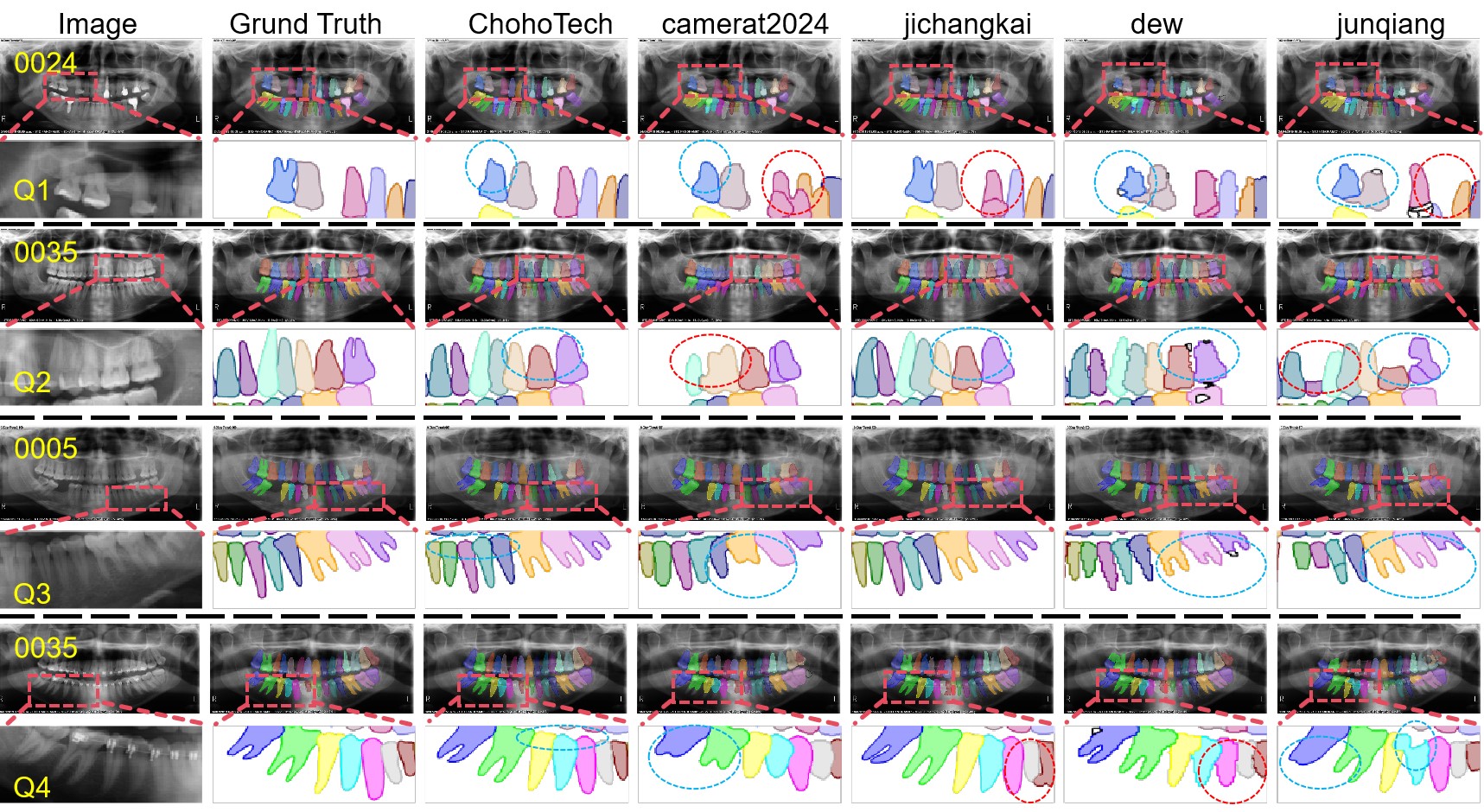}
    \caption{Qualitative comparison of 2D instance segmentation results from top-performing teams on representative adult panoramic X-ray (PXI) cases. Each case is presented with both a full view and a corresponding magnified quadrant view to facilitate detailed analysis of specific failure modes. These detailed views reveal common challenges, including the inaccurate segmentation of tooth apices (circled in \textcolor{blue}{blue}) and the erroneous merging of overlapping teeth into a single instance (circled in \textcolor{red}{red}).}
    \label{fig:2DPXI_adult}
\end{figure*}
\begin{figure*}[!ht]
    \centering
    \includegraphics[width=\linewidth]{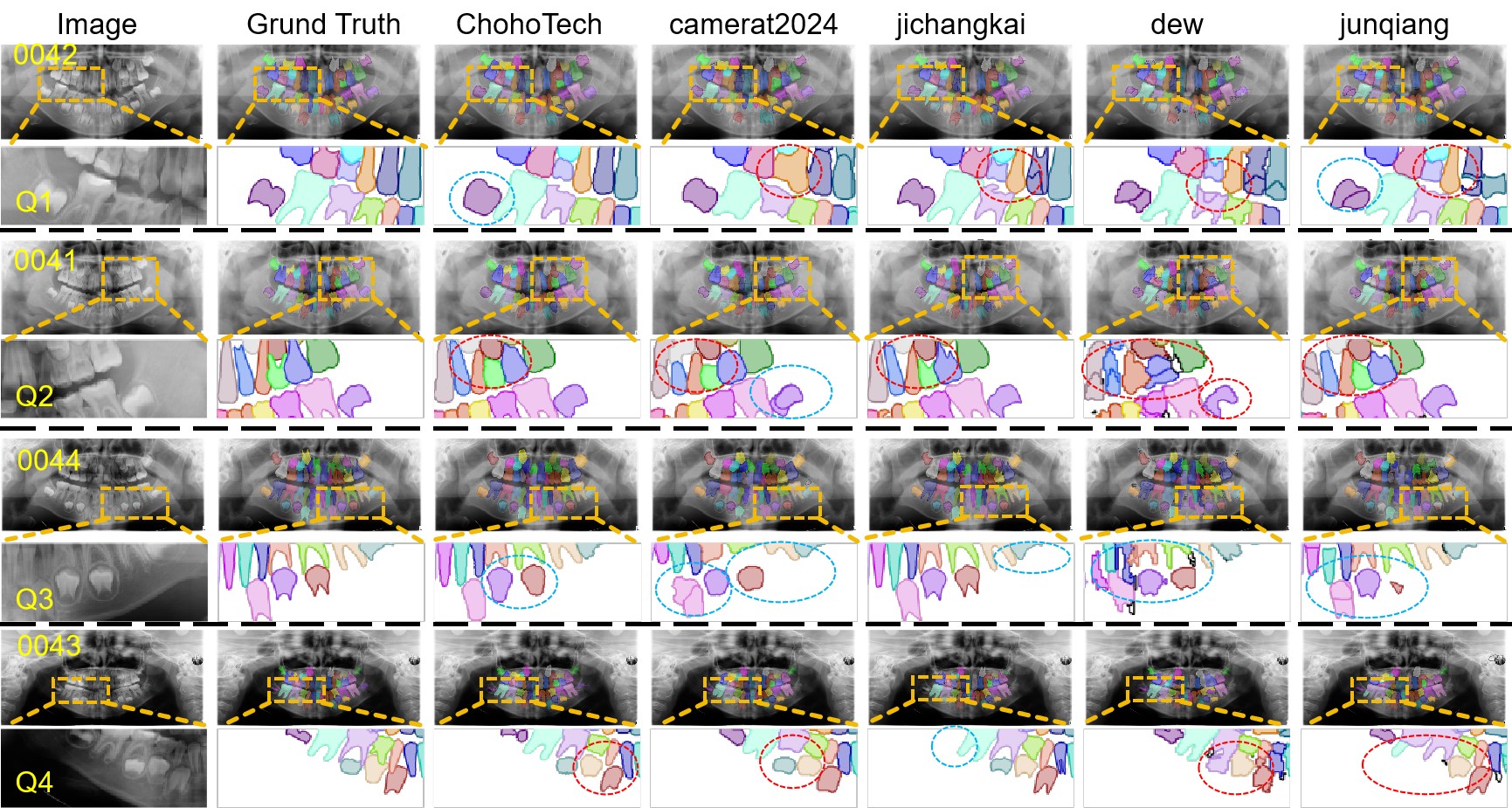}
    \caption{Qualitative comparison of 2D instance segmentation results from top-performing teams on representative pediatric panoramic X-ray (PXI) cases. Each case is presented with both a full view and a magnified quadrant view. Unlike adult dentition, pediatric cases present unique challenges due to the complex and irregular distribution of smaller teeth, leading to frequent and severe overlap. The magnified views highlight that most methods struggle in these dense, overlapping regions, often failing to accurately delineate individual small teeth, such as the erroneous merging of teeth (circled in \textcolor{red}{red}) or the inaccurate segmentation of developing apices (circled in \textcolor{blue}{blue}).}
    \label{fig:2DPXI_child}
\end{figure*}

\begin{figure*}[!ht]
    \centering
    \includegraphics[width=\linewidth]{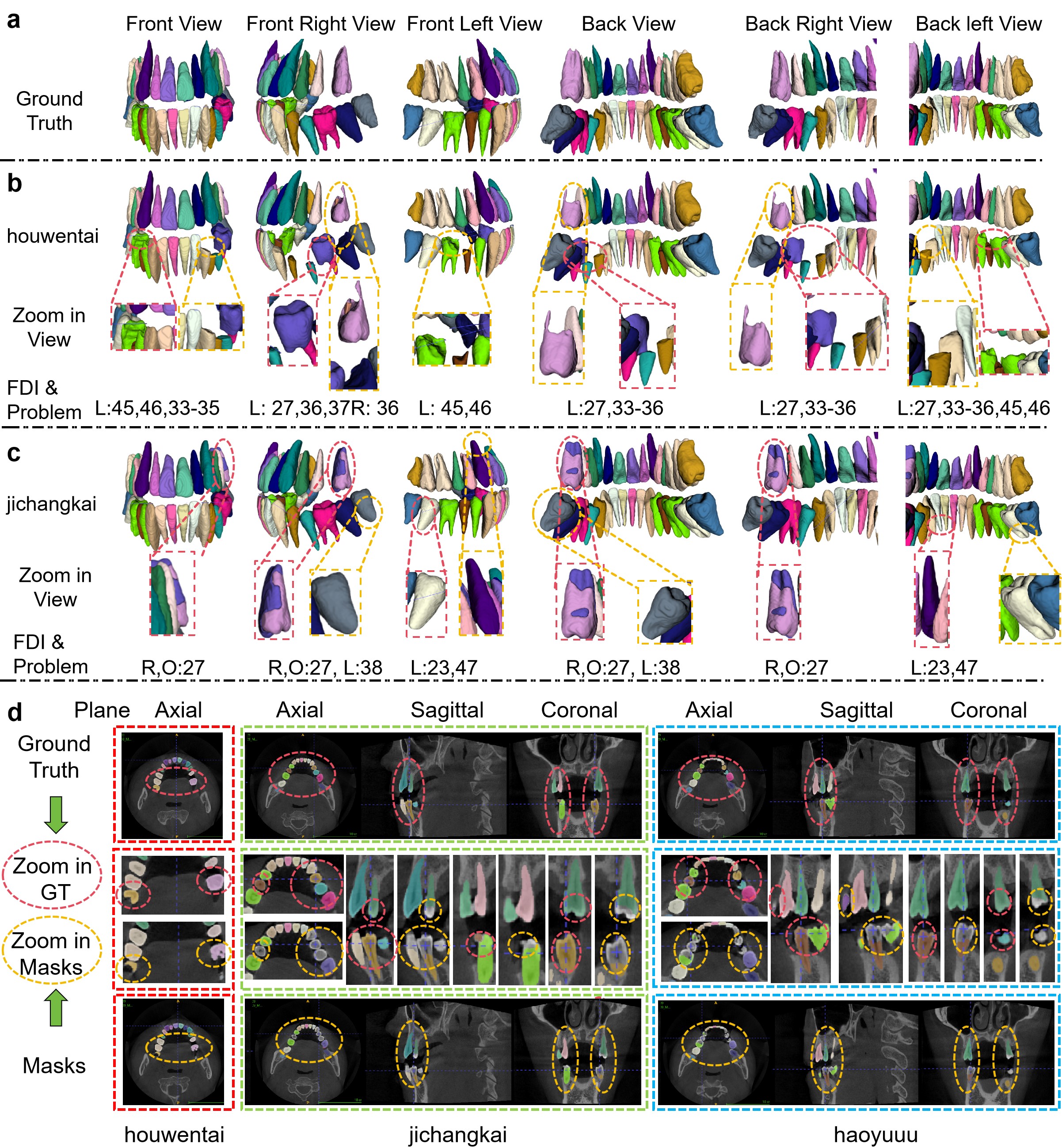}
    \caption{Visual comparison of 3D CBCT segmentation results. Multi-perspective views of the ground truth (a) are contrasted with outputs from top teams Houwentai and Jichangkai (b, c). Specific failure modes, including lack of segmentation (L), over-segmentation (O), and recognition errors (R) are indicated. A 2D slice comparison is provided in (d) for boundary detail.}
    \label{fig:3DCBCT_3dview}
\end{figure*}

\subsection{Qualitative Analysis of Segmentation Failures and Successes}

To complement the quantitative metrics, a qualitative analysis of the segmentation results provides crucial insights into the specific clinical and anatomical challenges that current semi-supervised methods face.

\subsubsection{Analysis of 2D PXI Segmentation}
Visual inspection of adult cases in Fig. \ref{fig:2DPXI_adult} reveals that while top-performing methods were proficient in segmenting well-defined teeth, two primary failure modes emerged in challenging regions. First, the erroneous merging of overlapping teeth (circled in red) was a common issue. In 2D panoramic projections, the boundaries between crowded or superimposed teeth become ambiguous. This suggests that SSL models relying heavily on local pixel context can struggle to enforce instance separation without a strong prior, a problem that detection-first architectures are designed to mitigate. Second, many methods produced inaccurate segmentations of tooth apices (circled in blue). The apex is a small, low-contrast structure of high clinical importance for diagnosing periapical lesions. Its poor segmentation likely indicates that standard region-based loss functions (e.g., Dice) are dominated by the larger tooth crown, thus failing to penalize errors on these small but critical anatomical targets.

These challenges were significantly amplified in the pediatric cases shown in Fig. \ref{fig:2DPXI_child}. Pediatric dentition, with its mix of smaller deciduous teeth and developing permanent teeth, creates a dense and highly irregular environment. The magnified views demonstrate that most methods struggled in these cluttered regions, often failing to delineate individual small teeth. This highlights a key difficulty for SSL: if the limited labeled data does not sufficiently represent the vast anatomical variability of pediatric development, the model may generate noisy or incomplete pseudo-labels for unlabeled pediatric cases, hindering its ability to learn robust features for these challenging objects.


    
\subsubsection{Qualitative Analysis of 3D Tracks at zoomed in view}

For the 3D track, Fig. \ref{fig:3DCBCT_3dview} illustrates typical error modes in volumetric segmentation. While leading methods successfully reconstructed the 3D morphology of most teeth, three primary failure patterns emerged: under-segmentation (e.g., merging adjacent teeth, labeled 'L'), over-segmentation (e.g., erroneous extensions into surrounding bone, labeled 'O'), and recognition errors (e.g., correct segmentation but incorrect tooth identification, labeled 'R'). These qualitative observations corroborate the quantitative findings, underscoring that accurate instance separation and identification remain critical challenges in 3D CBCT segmentation.





\begin{figure}[!ht]
    \centering
    \includegraphics[width=\linewidth]{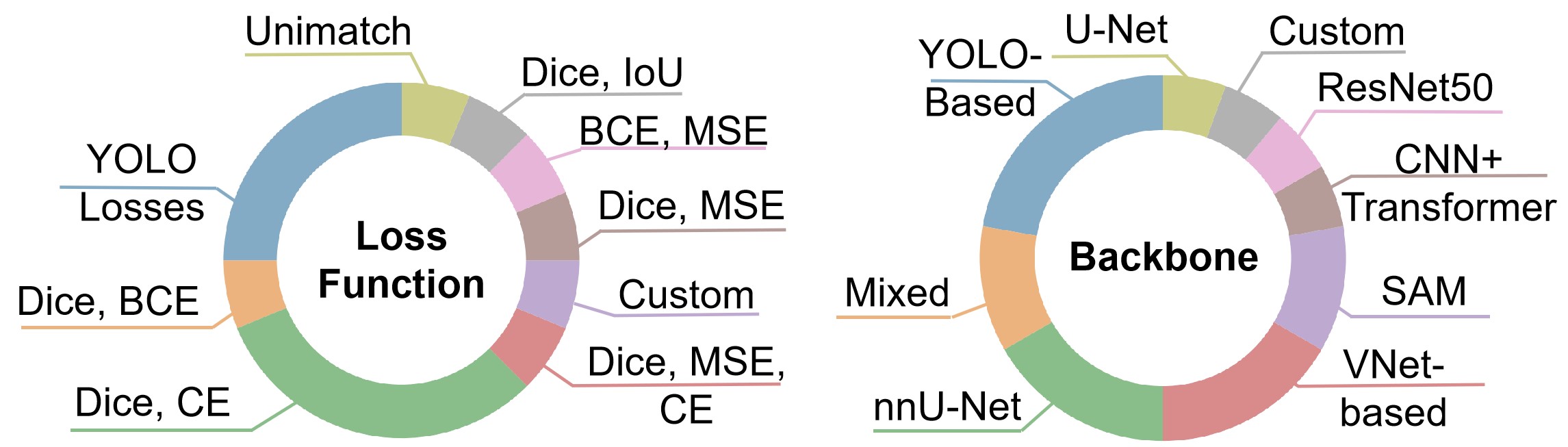}
    \caption{Distribution of loss functions employed and backbone architectures by participants in the 2D and 3D challenge tracks. (Left) Analysis of loss function usage indicates a strong preference for composite loss functions over single-objective training. The combination of a region-based loss (Dice) and a pixel-wise loss (Cross-Entropy) was the most common strategy. (Right) Breakdown of backbone model choices, revealing that nnU-Net and V-Net were the most prevalent architectures. A notable trend was the integration of the Segment Anything Model (SAM) as a foundational component. }
    \label{fig:backboneandloss}
\end{figure}

\subsection{Methodological Insights from Leading Teams}
An analysis of the methods submitted by the top-performing teams in both the 2D and 3D tracks reveals a clear consensus on several effective strategies, from architectural choices to training paradigms. While a diversity of approaches was observed, the most successful solutions consistently converged on a few key principles that effectively addressed the challenges of semi-supervised instance-level tooth segmentation. The distribution is depicted in Fig. \ref{fig:backboneandloss}.

\subsubsection{Network Architectures}
A dominant trend was the adoption of robust, highly specialized backbone architectures. The nnU-Net framework, renowned for its automated pipeline configuration and strong performance in medical imaging segmentation, emerged as a cornerstone for several leading teams, including Jichangkai and Houwentai. Its prevalence underscores the value of a well-tuned, powerful baseline in competitive challenges. Concurrently, some teams, such as ChohoTech and Caiyichen, successfully employed detection-based models like YOLOv8 and YOLOv9 as a foundational step. This indicates a strategic preference for two-stage "detect-then-segment" pipelines, where initial tooth localization simplifies the subsequent fine-grained segmentation task. Classic architectures like V-Net and U-Net also remained relevant, often adapted for specific tasks.

\subsubsection{Semi-supervised Techniques}
Beyond the choice of backbone, a key determinant of success was the implementation of multi-stage, coarse-to-fine refinement strategies. Rather than relying on a single end-to-end model, top teams often decomposed the complex problem into more manageable sub-tasks. The YOLO-based approaches are a prime example, using object detection for coarse localization before a dedicated segmentation head provides the final mask. Similarly, the winning method from Jichangkai implemented a two-stage process that first divided the image into quadrants and then performed high-resolution segmentation within each, effectively managing complexity and improving precision. This pattern suggests that explicitly guiding the model from a high-level overview to fine-grained detail is a highly effective strategy for dental instance segmentation.

\subsubsection{Loss Function}
Regarding the training objectives, no single loss function was universally dominant; instead, the clear trend was the use of composite loss functions. The most common combination for segmentation-focused models was a weighted sum of a region-based loss, typically the Dice loss, and a pixel-wise classification loss, such as Cross-Entropy (CE) or Binary Cross-Entropy (BCE). This hybrid approach balances the need for accurate spatial overlap (Dice) with correct pixel-level classification (CE/BCE), a standard practice that proved effective here. Teams employing YOLO-based backbones naturally utilized their corresponding specialized loss functions, such as CIoU, DFL, and VFL, tailored for detection and localization accuracy.

\subsubsection{Large-scale Foundational Models}
Finally, a notable emerging trend was the integration of large-scale foundational models, particularly the Segment Anything Model (SAM) and its medical-specific variant, SAM-Med2D. Teams like Isjinghao and Guo7777 leveraged these models, likely to generate high-quality pseudo-labels, provide strong feature initialization, or act as a powerful component within their segmentation pipeline. This highlights a strategic shift towards harnessing the powerful prior knowledge embedded in foundation models to enhance performance in data-scarce, semi-supervised scenarios. In summary, the leading solutions in the STS 2024 challenge skillfully combined robust backbones with multi-stage refinement pipelines, optimized them with composite loss functions, and, in several cases, augmented their performance by integrating knowledge from foundational models.

\subsubsection{Analysis of Limitations and Failed Approaches}
In contrast to the success of multi-stage and foundation-model-based approaches, an analysis of lower-performing entries reveals that the lack of explicit anatomical priors was a primary cause of failure. Methods relying solely on pixel-level consistency regularization often failed to capture the global topological arrangement of the dentition.
Specifically, in 2D OPG tasks, the absence of 'horizontal' spatial constraints (i.e., the relative ordering and separation of teeth along the dental arch) frequently led to the merging of adjacent, overlapping teeth. This failure mode was effectively mitigated in top-tier methods by using detection-based priors (e.g., YOLO bounding boxes) that enforce instance separation before segmentation. Similarly, in the 3D CBCT track, approaches that processed data without sufficient 'vertical' inter-slice consistency struggled to maintain volumetric integrity. Models lacking 3D context often produced fragmented segmentations, particularly for fine structures like tooth apices, as they failed to leverage the continuity of tooth anatomy across the Z-axis.
Furthermore, standard self-training pipelines without robust noise-filtering mechanisms (such as uncertainty estimation or ensemble voting) tended to overfit noisy pseudo-labels. This was particularly evident in 'hard' cases involving tooth loss or implants, where the domain shift in the unlabeled data led to error propagation, limiting the models' generalization capability."

\subsection{Performance Across Anatomical Regions and Demographics}
To further dissect performance, we analyzed Instance DSC scores in different anatomical quadrants and demographic groups, as shown in Tables~\ref{table:results_2d} and ~\ref{table:results_3d}.
For the 2D PXI track, the results were stratified by quadrant and by patient age (adult vs. child). This detailed breakdown confirms the observations from the qualitative analysis: on average, segmentation performance on pediatric cases was slightly lower than on adult cases across most teams and quadrants. For example, the top team, Chohotech achieved a total Instance DSC of over 82\%, with relatively stable performance across both adult and pediatric groups, demonstrating the robustness of their semi-supervised approach. This result is significant, as it confirms that SSL can be effectively applied to challenging pediatric data where labeled examples are particularly scarce, thereby addressing a key goal of this challenge.
In the 3D CBCT track, performance across quadrants was generally high for the top teams, with Houwentai and Chohotech achieving Instance DSC scores above 93\% in all regions. The lower-ranked teams showed more variability, indicating less consistent performance across the full dental arch.

\subsection{Algorithm Efficiency and Clinical Relevance}
The clinical applicability of AI models depends not only on accuracy but also on computational efficiency. Table~\ref{table:results_2d} and ~\ref{table:results_3d} provides metrics for GPU memory consumption (AUC\_GPU) and runtime (RT).
In the 2D track, the top-ranked method from ChohoTech was not only accurate but also efficient, with a runtime of approximately 13 seconds per image. Several other methods also achieved inference times under 20 seconds, suggesting a strong potential for integration into clinical workflows where a near-real-time response is desirable. In contrast, team Jichangkai's method, while highly accurate, required a significantly longer runtime (55.90s), which might be less suitable for interactive applications.
The computational demands for the 3D track were substantially higher. The fastest method (Madongdong at 52.82s) was still considerably slower than the 2D methods. Other top-performing 3D models required several minutes for inference, and their GPU memory consumption was an order of magnitude higher than their 2D counterparts. This starkly illustrates the trade-off between the detailed anatomical information provided by 3D CBCT and the computational resources required to process it, a critical consideration for deployment in real-world clinical settings.


\section{Discussion}

\subsection{Main findings}
Dental image segmentation presents unique challenges that differentiate it from other medical imaging domains. The STS 2024 Challenge underscored several critical difficulties inherent to dental datasets: 1) Limited Segmentation Accuracy of Previous Methods; 2) Inadequate Integration of Segmentation and Tooth Position Recognition; 3) High Proportion of Unannotated Data. Traditional methods rely heavily on fully supervised learning paradigms, which demand extensively annotated datasets. However, annotating dental images is labor-intensive and requires specialized expertise, leading to a scarcity of high-quality labeled data. Consequently, previous approaches exhibited limited precision in accurately delineating tooth boundaries, especially in cases involving overlapping structures or varying image qualities. The complexity of dental anatomy, with its intricate and closely packed tooth structures, further exacerbates segmentation inaccuracies.

\hl{Beyond mere segmentation,} accurate tooth position identification is crucial for comprehensive dental analysis and clinical decision-making. Prior methodologies typically treated segmentation and tooth position recognition as separate tasks, leading to suboptimal performance in integrated scenarios. This discrimination fails to leverage the inter-dependencies between segmentation and positional data, resulting in inconsistencies and reduced overall accuracy. The inability to concurrently address segmentation and tooth position recognition limits the utility of automated systems in practical dental applications.

\hl{The STS 2024 Challenge introduced several innovative elements in dataset construction and challenge design to address the aforementioned challenges. }First, to mitigate the scarcity of annotated data, we meticulously curated a dataset encompassing 2D panoramic X-ray images and 3D CBCT tooth volumes. This multiple-modal dataset enriches the dataset and facilitates the development of algorithms capable of handling different imaging modalities, encouraging the creation of more robust and generalizable models.

\hl{ Furthermore,} including pediatric and adult dental images ensures that models can adapt to variations across age groups, enhancing their applicability in real-world clinical settings. Our challenge incorporated a multi-faceted evaluation framework that assesses participants' algorithms on several levels, ensuring a comprehensive performance evaluation: Instance-Level Evaluation, Image-Level Evaluation.

The results from the STS 2024 Challenge reveal several noteworthy insights: Top-performing teams predominantly leveraged semi-supervised learning frameworks, effectively utilizing the limited labeled data alongside the abundant unlabeled data. Techniques such as integrating pre-trained models like the Segment Anything Model (SAM), consistency regularization learning, and multi-stage architecture optimization significantly enhanced segmentation accuracy.\hl{This trend aligns with the broader trend in medical imaging}, where semi-supervised methods are increasingly recognized for their ability to overcome data scarcity.

\hl{In the 3D CBCT segmentation task,} multi-stage training strategies yielded superior performance. Participants could incrementally improve segmentation precision and robustness by sequentially refining the model through multiple training phases. This approach facilitates better feature learning and mitigates the risk of overfitting, particularly in complex 3D structures. 

\hl{Despite the overall success of semi-supervised methods,} a notable portion of participants did not employ any semi-supervised strategies. This variability reveals a critical gap between the demonstrated potential of SSL and its practical adoption, which have demonstrated clear advantages in handling unannotated data. 

\hl{Across the diverse methodologies, }participants predominantly introduced perturbations at multiple levels to augment model training and generalization: Data-Level Perturbations: Techniques such as weak-strong augmentations and the addition of Gaussian noise were employed to create varied input data representations, fostering robustness against input variations. Model-Level Perturbations: The integration of heterogeneous network architectures (e.g., CNNs and Transformers) facilitated implicit consistency regularization, promoting diverse feature learning and reducing model-specific biases. Training Cycle Perturbations: Frameworks like the teacher-student model incorporated temporal perturbations by iteratively refining pseudo-labels, thereby enhancing feature learning and mitigating overfitting. 

\hl{In terms of consistency constraints}, methods varied across three primary dimensions: Soft Constraints: Utilization of soft logits and features as supervisory signals encouraged smooth decision boundaries and feature consistency across different perturbations. Hard Constraints: Binary segmentation outputs were enforced through hard label assignments, ensuring definitive delineation of anatomical structures. Structural Constraints: Incorporation of edge detection operators (e.g., Sobel) imposed structural integrity on anatomical boundaries, aligning segmentation outputs with inherent anatomical features.

Building upon the insights gained from the STS 2024 Challenge, several avenues for future research and development emerge. First, efforts should be directed toward creating more extensive and diverse annotated dental datasets encompassing various anatomical variations, imaging modalities, and clinical conditions. Collaborative initiatives and data-sharing platforms can facilitate the accumulation of comprehensive datasets necessary for training robust deep-learning models. 

\hl{Further exploration of semi-supervised and unsupervised learning methods can enhance model performance and reduce dependency on labeled data.} Techniques such as self-supervised learning, transfer learning, and active learning hold promise for improving segmentation accuracy and efficiency. Developing models that concurrently address segmentation and tooth position identification can yield more holistic and clinically relevant outputs. Multi-task learning frameworks can enhance the synergy between tasks, improving overall performance and utility. Bridging the gap between research and clinical practice requires rigorous validation of segmentation models in real-world settings. Future studies should focus on integrating automated segmentation tools into clinical workflows and assessing their impact on diagnostic accuracy, treatment planning, and patient outcomes.

\hl{Enhancing the computational efficiency of semi-supervised models is crucial for their adoption in clinical environments.} Research should optimize models for real-time processing without compromising segmentation accuracy, ensuring their practicality and scalability. Subsequent iterations of the STS Challenge can incorporate more diverse and complex datasets, introduce additional evaluation metrics that reflect clinical utility, and encourage the development of lightweight models suitable for deployment in various clinical settings. Such enhancements can further drive innovation and accelerate the adoption of advanced segmentation techniques in dentistry.

\begin{figure*}[!ht]
    \centering
    \includegraphics[width=\linewidth]{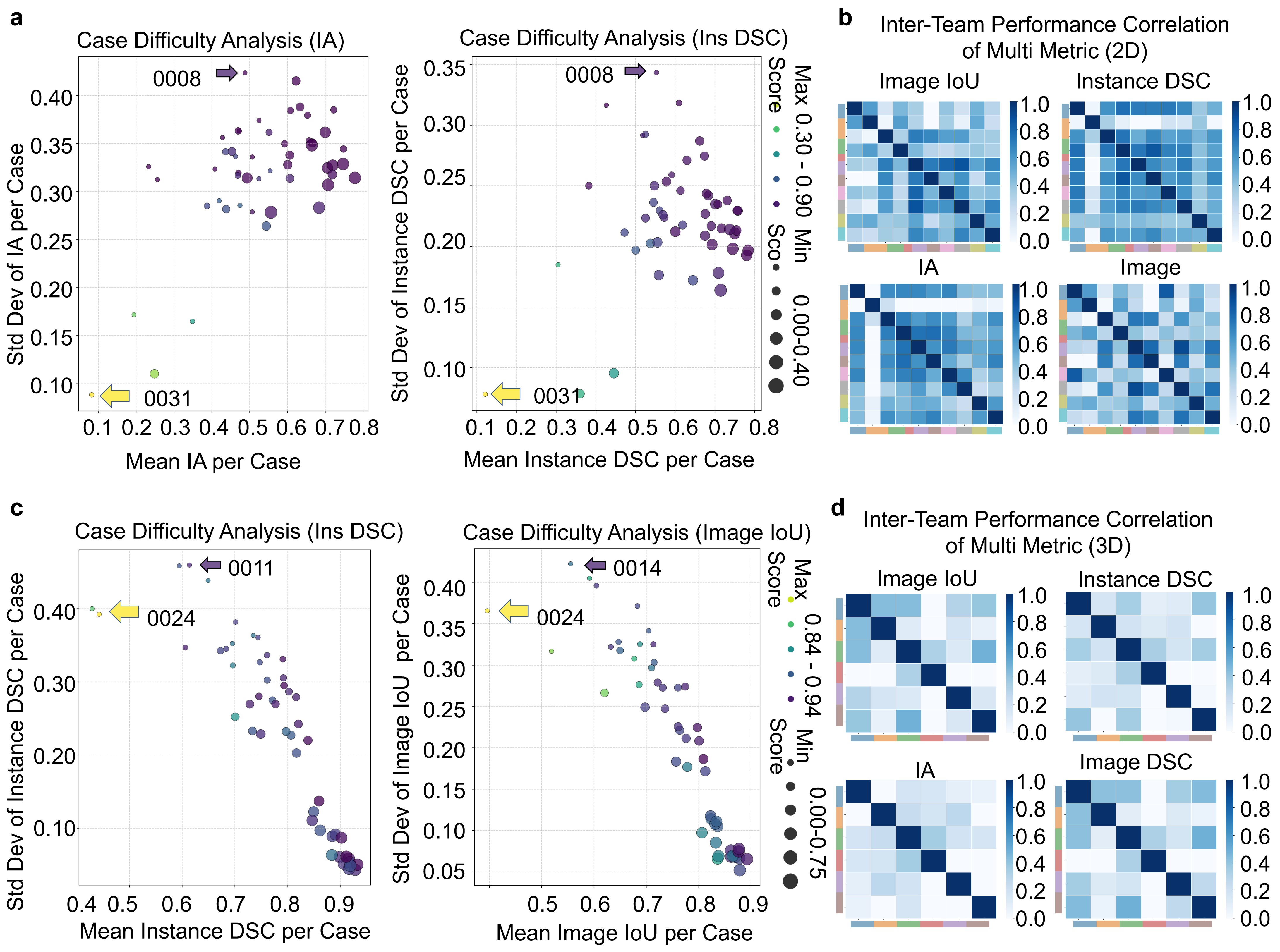}
    \caption{Case difficulty and inter-metric correlation analysis for the 2D (a, b) and 3D (c, d) tracks.
(a, c) Case Difficulty Analysis: Plots identifying the most challenging cases (0031 in 2D, 0024 in 3D). Difficulty in 2D was linked to severe tooth loss, while in 3D, it arose from small targets (e.g., apices) and inter-tooth misidentification.
(b, d) Inter-Metric Correlation: Matrices for both tracks showing strong concordance across metrics, indicating consistent team rankings.}
    \label{fig:relation_2d_01}
\end{figure*}

\subsection{Case Difficulty and Metric Correlation}

The case difficulty and metric correlation analysis in Fig.~\ref {fig:relation_2d_01} offers valuable insights into the dataset's intrinsic challenges and the robustness of our evaluation framework.

The difficulty analysis in Fig.~\ref {fig:relation_2d_01}(a, c) reveals that the primary challenges differed by modality. In the 2D PXI track, the most difficult cases (e.g., Case 0031, which had the lowest scores)  were those with severe pathologies like extensive tooth loss. This is a classic problem for semi-supervised learning, as these atypical cases deviate from the primary distribution of the unlabeled data, hindering the generation of effective pseudo-labels. This suggests a need for SSL methods that are more robust to domain shift within the dataset itself. \hl{In contrast, 3D CBCT challenges were often related to acquisition variations and fine anatomical details. For instance, Case 0024 was identified as the most challenging 3D sample. Visual inspection revealed that this volume suffered from a severe scan position offset, where the dentition was shifted entirely to the anterior third of the image space. This spatial anomaly caused failure in models relying on standard centering priors.} This points to architectural limitations in current models, which may require more specialized designs to capture fine-grained features in volumetric data.

The strong positive correlation between different evaluation metrics, as shown in the matrices from Fig. ~\ref {fig:relation_2d_01}(b, d), is an interesting finding. The high concordance across metrics like Instance DSC and Instance Affinity (IA) indicates that the team rankings were stable and not dependent on a specific metric choice. This validates the robustness of our evaluation framework and suggests that the top-performing methods were genuinely superior across multiple criteria of segmentation quality.

\begin{figure*}[!ht]
    \centering
    \includegraphics[width=\linewidth]{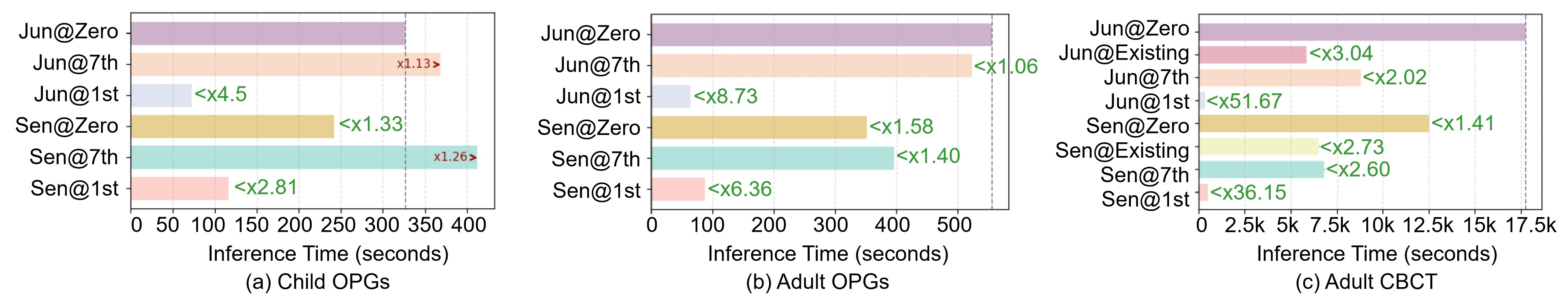}
    \caption{Bar plots of X-ray and CBCT labeling time under varied methods. The \textcolor{green}{green arrow} indicates a time reduction (acceleration) and the \textcolor{red}{red arrow} an increase (deceleration) compared to the baseline (typically the Junior@Zero). Jun: Junior dentist; Sen: Senior dentist.}
    \label{fig:acc_rates}
\end{figure*}

\subsection{Human-Machine Collaborative Annotation Study}
To quantitatively evaluate the clinical utility of semi-supervised segmentation algorithms in this challenge, we conducted a human-machine collaborative annotation study. The study involved two dentists (a junior clinician with more than two years of experience and a senior specialist with more than ten years of experience). \hl{The dentists annotated a curated test set of 30 dental scans, including 10 X-rays from child, 10 X-rays images from adult, and 10 CBCT volumes from adult. To simulate a realistic clinical workflow, dentists utilized LabelMe for 2D X-ray polygon annotation and 3D Slicer for 3D CBCT volumetric refinement. These industry-standard tools were used across all strategies to ensure consistent time measurement.}  The study follows four distinct strategies:
(i) Full manual annotation (i.e, \textit{From scratch});
(ii) Correction of predictions from existing methods (only available for the CBCT);
(iii) Correction of $\text{7}^\text{th}$-ranked outputs (relatively low-quality); and
(iv) Correction of $\text{1}^\text{st}$-ranked outputs (relatively high-quality).

Annotation time per case was rigorously recorded to assess workflow efficiency, as shown in Fig.~\ref{fig:acc_rates}. This study focuses on three key dimensions: (i) the relationship between algorithmic performance (as reflected by model rankings) and time saved during clinical workflows; (ii) modality-specific challenges in 2D vs. 3D tasks, including anatomical variability across pediatric and adult populations; and (iii) how clinician expertise interacts with automated tools to optimize workflow efficiency.

\subsubsection{Dramatic Time Savings in Clinical Workflows}
The top-performing semi-supervised model (the $\text{1}^\text{st}$-ranked team) demonstrated transformative efficiency gains across all modalities. In 2D X-rays (a historically unaddressed domain), the labeling efficiency improved 4.5$\times$ for the junior dentists (i.e., from 326.10s to 72.50s per case) and 2.11$\times$ for the senior (i.e., from 244.50s to 116.10s). Critically, the $\text{1}^\text{st}$-ranked model outperformed existing CBCT methods by 94.1\% (i.e., the junior's time that decreased from 5841.80s to 343.30s), validating its superiority in handling sparse-annotation scenarios. For the CBCT volumes, the annotation time plummeted from 3.5 \~ 4.9 hours to 5.7 \~ 8.2 minutes per volume, which is clinically critical for efficient orthodontics and implant planning.

Moreover, the segmentation quality directly impacts the annotation efficiency. The relatively low-quality initial predictions may increase the cognitive load and correction time of dentists. For instance, the $\text{7}^\text{th}$-ranked model prolonged senior dentists’ child X-ray annotation by 68.5\% (i.e., increasing from 244.50s to 412.00s), indicating that subpar segmentation disrupts clinical efficiency. Conversely, the $\text{1}^\text{st}$-ranked model reduced time while improving consistency, where the standard deviations decreased across tasks. This highlights that only high-precision algorithms enable reliable human-machine collaboration, which is one of the core goal of our semi-supervised challenge.

Notably, the CBCT modality that is essential for complex orthodontic procedures, showed the greatest absolute time savings (e.g., from 4.9 hours to $\leq$0.1 hour / volume). This aligns with our challenge’s focus on underrepresented 3D data, addressing a critical barrier in dental area. For adult X-rays, the $\text{1}^\text{st}$-ranked model reduced annotation time by 88.2\% (e.g., the junior's time decreasd from 555.50s to 63.60s), outperforming the $\text{7}^\text{th}$-place model by 87.7\%. Notably, for pediatric X-rays—a domain previously lacking public datasets—the methods achieved near-adult levels of efficiency (e.g., 72.5s vs. 65.7s for the junior), proving semi-supervised methods can overcome pediatric data scarcity.

\subsubsection{Implications for Clinical Deployment}
The junior dentist benefited disproportionately from high-quality AI assistance. In pediatric X-rays, their efficiency gain with the $\text{1}^\text{st}$-ranked model (4.5$\times$) exceeded Seniors’ (2.1$\times$), narrowing the experience gap: the junior-senior time differential dropped from 81.6s (manual) to 43.6s (AI-assisted). However, Seniors exhibited lower tolerance for imperfect predictions—when using the $\text{7}^\text{th}$-place model, their pediatric X-ray annotation time surged by 68.5\%, versus only 12.9\% for Juniors. This implies that AI assistance most empowers early-career clinicians, while experts require near-perfect segmentation for workflow integration. \hl{Based on the measured throughput (decreasing from ~ 4 hours to ~5 minutes per volume), we estimate that a Senior dentist could process approximately 96 CBCT volumes/8-hour workday with 1st-ranked model assistance versus 2 volumes/workday manually—enabling large-scale data curation for AI development.} Combining Junior clinicians with state-of-the-art AI achieved 73\% of Senior-level efficiency (e.g., pediatric X-rays: 72.5s vs. Senior’s 116.1s) at lower resource cost. This validates our challenge’s design premise: semi-supervised learning bridges data scarcity in specialized domains (pediatrics, CBCT) while optimizing clinical resource utilization.

In a nutshell, our challenge’s top semi-supervised model reduced dental annotation effort by 52–98\%, demonstrating that human-AI collaboration—when powered by high-precision initial segmentation—can overcome historic barriers in pediatric and 3D dental imaging, accelerating the translation of AI tools into clinical practice. Our challenge catalyzed algorithms that transform dental image annotation from a hours-long manual task to a minutes-long collaborative effort, with profound implications for clinical scalability and global dental health equity. Future work should focus on real-time interactive refinement tools to further harness human-AI synergies. The results underscore the critical role of semi-supervised algorithms in enabling scalable, cost-effective dental diagnostics while maintaining clinical accuracy—a prerequisite for widespread adoption in resource-constrained settings. Also, this study directly addresses two critical gaps in dental AI research: (1) the absence of benchmarks for human-AI collaboration efficiency in pediatric dental imaging, and (2) the lack of comparative data on how semi-supervised model quality impacts real-world clinical labor. By simulating real-world refinement scenarios, we quantify how algorithmic performance translates to tangible clinical time savings and evaluate the viability of semi-supervised solutions for underserved tasks (e.g., pediatric X-rays and CBCT segmentation, where public datasets are absent).

\begin{hlbox}
\subsection{Limitations and Future Directions}
While the STS 2024 Challenge made strides in advancing dental image segmentation, it has limitations. The dataset, although diverse, remains limited in size and scope, potentially affecting the generalizability of the findings. Additionally, the challenge primarily focused on segmentation accuracy, with less emphasis on other critical factors such as processing speed and integration with clinical workflows. Furthermore, while representative of real-world scenarios, the high proportion of unannotated data may have introduced biases that could influence the performance of semi-supervised algorithms. 

Looking ahead, the trajectory of this field must shift from algorithmic competition to clinical translation. A critical open question is whether current performance levels are sufficient to support high-precision autonomous systems, such as AI-guided robotic dental implant surgery, where error margins are measured in fractions of a millimeter. To answer this, future research should move beyond geometric metrics (like Dice) and incorporate downstream functional benchmarks—evaluating, for instance, how segmentation errors impact the stability of orthodontic movement simulations or the fit of 3D-printed surgical guides.
Furthermore, investigating the theoretical 'upper bound' of performance with fully labeled, multi-expert consensus datasets is essential to quantify the exact efficiency gap of SSL methods. Finally, to ensure these tools are safe for global deployment, future initiatives must prioritize large-scale, multi-center clinical validation to test model robustness across heterogeneous scanner protocols and diverse patient demographics, ultimately bridging the divide between academic benchmarks and reliable clinical practice.
\end{hlbox}

\section{Conclusion}

The STS 2024 Challenge represents an advancement in semi-supervised instance segmentation for dental imaging, establishing the first standardized benchmark for label-efficient learning in this clinically critical domain. Our comprehensive analysis reveals that successful methodologies consistently leveraged synergistic combinations of three core strategies: knowledge transfer from foundation models, iterative pseudo-label refinement, consistency regularization leanring, and multi-stage architectural optimization. The dominance of teacher-student frameworks employed by 85\% of top-ranked solutions underscores their effectiveness in propagating knowledge from limited labeled data through carefully designed pseudo-labeling cycles. Notably, the integration of Segment Anything Model (SAM) variants and nnU-Net backbones emerged as particularly potent, enabling teams like Guo777 and Isjinhao to achieve instance-level Dice scores exceeding 90\% in CBCT segmentation while utilizing merely 9\% labeled data. These technical innovations directly translate to tangible clinical value, as evidenced by our human-machine collaboration study demonstrating that top solutions reduced annotation time by 88.2\% for adult OPG and 94.1\% for CBCT volumes when compared to manual annotation with junior clinicians achieving 73\% of senior-expert efficiency when refining high-quality predictions.

Despite these advances, several critical challenges persist. The underutilization of unlabeled data by approximately 60\% of participating teams highlights a persistent gap between semi-supervised learning potential and practical implementation. Furthermore, while multi-stage pipelines like Haoyuuu's T3Net showed exceptional performance in 3D segmentation, their computational complexity presents deployment barriers in time-sensitive clinical settings. Limitations in dataset diversity, particularly the scarcity of pediatric CBCT scans and complex pathological cases, also constrain model generalizability. Looking forward, three priority directions emerge: 1) Developing hybrid SSL frameworks that combine SAM's zero-shot generalization with uncertainty-aware consistency constraints, 2) Creating cross-modal learning architectures that leverage complementary information from both 2D and 3D dental imaging, and 3) Establishing standardized clinical validation protocols that assess both segmentation accuracy and real-world workflow integration. The publicly released STS 2024 dataset and open-sourced methodologies provide an essential foundation for addressing these priorities. By catalyzing research in label-efficient learning for dental instance segmentation, this initiative accelerates progress toward accessible, AI-enhanced diagnostics that can transform global oral healthcare, particularly in resource-constrained settings where expert annotation remains scarce.

\section*{CRediT authorship contribution statement}


\textbf{Yaqi Wang}: Conceptualization, Investigation, Funding acquisition
\textbf{Jun Liu}: Supervision, Project administration, Conceptualization, Funding acquisition, Writing – Review \& Editing
\textbf{Yifan Zhang}: Data Curation, Resources
\textbf{Zhi Li}: Writing–Review \& Editing, Validation, Visualization
\textbf{Jiaxue Ni}: Investigation, Validation
\textbf{Qian Luo}: Writing-Review, Validation
\textbf{Jialuo Chen}: Writing-Review, Validation
\textbf{Chengyu Wu}: Writing-Original Draft \& Editing, Validation
\textbf{Hongyuan Zhang}: Editing, Validation, Visualization
\textbf{Jin Liu}: Investigation, Formal analysis
\textbf{Can Han}: Investigation, Validation
\textbf{Kaiwen Fu}: Competitor
\textbf{Changkai Ji}: Competitor
\textbf{Xinxu Cai}: Competitor
\textbf{Jing Hao}: Competitor
\textbf{Zhihao Zheng}: Competitor
\textbf{Shi Xu}: Competitor
\textbf{Junqiang Chen}: Competitor
\textbf{Qianni Zhang}: Investigation, Formal analysis
\textbf{Dahong Qian}: Investigation, Validation
\textbf{Shuai Wang}: Supervision, Methodology, Software, Validation, Writing – Review \& Editing
\textbf{Huiyu Zhou}: Supervision, Project administration, Conceptualization, Review \& Editing

\section*{Declaration of competing interest}
The authors declare that they have no known competing financial interests or personal relationships that could have appeared to influence the work reported in this paper.

\section*{Acknowledgements}
This work was supported by the National Natural Science Foundation of China (No. 62206242, No. 62201323), Zhejiang Provincial Natural Science Foundation of China (No. LD25F020005), and China Science and Technology Foundation of Sichuan Province (No. 2022YFS0116). There are no conflicts of interest between authors. Yifan Zhang is the principal sponsor of the challenge by collecting and providing clinical data. Only the organizers and members of their immediate team have access to test case labels. The study protocol was approved by the Medical Ethics Committee of Hangzhou Stomatological Hospital (Approval No: 2022YR014).

\section*{Data availability}
Data will be made available on request.

\bibliographystyle{elsarticle-num} 
\bibliography{refs} 

@inproceedings{bolelli2025segmenting,
  title={Segmenting Maxillofacial Structures in CBCT Volumes},
  author={Bolelli, Federico and Marchesini, Kevin and van Nistelrooij, Niels and Lumetti, Luca and Pipoli, Vittorio and Ficarra, Elisa and Vinayahalingam, Shankeeth and Grana, Costantino},
  booktitle={Proceedings of the Computer Vision and Pattern Recognition Conference},
  pages={5238--5248},
  year={2025}
}

@article{bolelli2024segmenting,
  title={Segmenting the inferior alveolar canal in cbcts volumes: the toothfairy challenge},
  author={Bolelli, Federico and Lumetti, Luca and Vinayahalingam, Shankeeth and Di Bartolomeo, Mattia and Pellacani, Arrigo and Marchesini, Kevin and Van Nistelrooij, Niels and Van Lierop, Pieter and Xi, Tong and Liu, Yusheng and others},
  journal={IEEE Transactions on Medical Imaging},
  year={2024},
  publisher={IEEE}
}

@article{lumetti2024enhancing,
  title={Enhancing patch-based learning for the segmentation of the mandibular canal},
  author={Lumetti, Luca and Pipoli, Vittorio and Bolelli, Federico and Ficarra, Elisa and Grana, Costantino},
  journal={IEEE Access},
  volume={12},
  pages={79014--79024},
  year={2024},
  publisher={IEEE}
}

@article{sischo2011oral,
  title={Oral health-related quality of life: what, why, how, and future implications},
  author={Sischo, Lacey and Broder, HL21422477},
  journal={Journal of dental research},
  volume={90},
  number={11},
  pages={1264--1270},
  year={2011},
  publisher={SAGE Publications Sage CA: Los Angeles, CA}
}

@article{shah2014recent,
  title={Recent advances in imaging technologies in dentistry},
  author={Shah, Naseem and Bansal, Nikhil and Logani, Ajay},
  journal={World journal of radiology},
  volume={6},
  number={10},
  pages={794},
  year={2014}
}

@article{cosson2020interpreting,
  title={Interpreting an orthopantomogram},
  author={Cosson, John},
  journal={Australian Journal of General Practice},
  volume={49},
  number={9},
  pages={550--555},
  year={2020},
  publisher={Royal Australian College of General Practitioners Sydney}
}

@article{jain2019new,
  title={New evolution of cone-beam computed tomography in dentistry: Combining digital technologies},
  author={Jain, Supreet and Choudhary, Kartik and Nagi, Ravleen and Shukla, Stuti and Kaur, Navneet and Grover, Deepak},
  journal={Imaging science in dentistry},
  volume={49},
  number={3},
  pages={179},
  year={2019}
}

@article{xu20183d,
  title={3D tooth segmentation and labeling using deep convolutional neural networks},
  author={Xu, Xiaojie and Liu, Chang and Zheng, Youyi},
  journal={IEEE transactions on visualization and computer graphics},
  volume={25},
  number={7},
  pages={2336--2348},
  year={2018},
  publisher={IEEE}
}

@article{lahoud2021artificial,
  title={Artificial intelligence for fast and accurate 3-dimensional tooth segmentation on cone-beam computed tomography},
  author={Lahoud, Pierre and EzEldeen, Mostafa and Beznik, Thomas and Willems, Holger and Leite, Andr{\'e} and Van Gerven, Adriaan and Jacobs, Reinhilde},
  journal={Journal of Endodontics},
  volume={47},
  number={5},
  pages={827--835},
  year={2021},
  publisher={Elsevier}
}

@article{sapkota2024zero,
  title={Zero-shot automatic annotation and instance segmentation using llm-generated datasets: Eliminating field imaging and manual annotation for deep learning model development},
  author={Sapkota, Ranjan and Paudel, Achyut and Karkee, Manoj},
  journal={arXiv preprint arXiv:2411.11285},
  year={2024}
}

@article{qin2023flexssl,
  title={FlexSSL: A Generic and Efficient Framework for Semi-Supervised Learning},
  author={Qin, Huiling and Zhan, Xianyuan and Li, Yuanxun and Zheng, Yu},
  journal={arXiv preprint arXiv:2312.16892},
  year={2023}
}

@article{kumarihami2018development,
  title={Development of Brief Image Quality Evaluation Criteria for Digital OrthoPantomography (OPG) Images in Dental Radiography},
  author={Kumarihami, AMC and Heshani, SDL and Sathyathas, P and Illeperuma, RP},
  journal={Journal of Health Science},
  volume={6},
  pages={139--147},
  year={2018}
}

@article{jiang2023instance,
  title={Instance recognition of street trees from urban point clouds using a three-stage neural network},
  author={Jiang, Tengping and Wang, Yongjun and Liu, Shan and Zhang, Qinyu and Zhao, Lin and Sun, Jian},
  journal={ISPRS Journal of Photogrammetry and Remote Sensing},
  volume={199},
  pages={305--334},
  year={2023},
  publisher={Elsevier}
}

@article{wang2025miccai,
  title={MICCAI 2023 STS Challenge: A retrospective study of semi-supervised approaches for teeth segmentation},
  author={Wang, Yaqi and Zhang, Yifan and Chen, Xiaodiao and Wang, Shuai and Qian, Dahong and Ye, Fan and Xu, Feng and Zhang, Hongyuan and Dan, Ruilong and Zhang, Qianni and others},
  journal={Pattern Recognition},
  pages={112049},
  year={2025},
  publisher={Elsevier}
}

@article{panetta2021tufts,
  title={Tufts dental database: a multimodal panoramic x-ray dataset for benchmarking diagnostic systems},
  author={Panetta, Karen and Rajendran, Rahul and Ramesh, Aruna and Rao, Shishir Paramathma and Agaian, Sos},
  journal={IEEE journal of biomedical and health informatics},
  volume={26},
  number={4},
  pages={1650--1659},
  year={2021},
  publisher={IEEE}
}

@inproceedings{cui2022ctooth+,
  title={Ctooth+: A large-scale dental cone beam computed tomography dataset and benchmark for tooth volume segmentation},
  author={Cui, Weiwei and Wang, Yaqi and Li, Yilong and Song, Dan and Zuo, Xingyong and Wang, Jiaojiao and Zhang, Yifan and Zhou, Huiyu and Chong, Bung san and Zeng, Liaoyuan and others},
  booktitle={MICCAI Workshop on Data Augmentation, Labelling, and Imperfections},
  pages={64--73},
  year={2022},
  organization={Springer}
}

@article{wang2023multi,
  title={Multi-level uncertainty aware learning for semi-supervised dental panoramic caries segmentation},
  author={Wang, Xianyun and Gao, Sizhe and Jiang, Kaisheng and Zhang, Huicong and Wang, Linhong and Chen, Feng and Yu, Jun and Yang, Fan},
  journal={Neurocomputing},
  volume={540},
  pages={126208},
  year={2023},
  publisher={Elsevier}
}

@article{huang2024multimodal,
  title={A multimodal dental dataset facilitating machine learning research and clinic services},
  author={Huang, Yunyou and Liu, Wenjing and Yao, Caiqin and Miao, Xiuxia and Guan, Xianglong and Lu, Xiangjiang and Liang, Xiaoshuang and Ma, Li and Tang, Suqin and Zhang, Zhifei and others},
  journal={Scientific Data},
  volume={11},
  number={1},
  pages={1291},
  year={2024},
  publisher={Nature Publishing Group UK London}
}

@misc{budagam2024instance, title={Instance Segmentation and Teeth Classification in Panoramic X-rays}, author={Devichand Budagam and Ayush Kumar and Sayan Ghosh and Anuj Shrivastav and Azamat Zhanatuly Imanbayev and Iskander Rafailovich Akhmetov and Dmitrii Kaplun and Sergey Antonov and Artem Rychenkov and Gleb Cyganov and Aleksandr Sinitca}, year={2024}, eprint={2406.03747}, archivePrefix={arXiv}, primaryClass={cs.CV} }

@article{roman2021panoramic,
  title={Panoramic Dental Radiography Image Enhancement Using Multiscale Mathematical Morphology},
  author={Rom{\'a}n, Juan Carlos Monta{\~n}a and Fretes, Victor R and Adorno, Carlos G and Silva, Rafael G and Noguera, Juan Luis Villalba and Legal-Ayala, Hugo and Mello-Rom{\'a}n, Juan D and Torres, Ra{\'u}l Dario Escobar and Facon, Jacques},
  journal={Sensors},
  volume={21},
  number={9},
  pages={3110},
  year={2021},
  publisher={MDPI},
  doi={10.3390/s21093110}
}

@inproceedings{redmon2016you,
  title={You only look once: Unified, real-time object detection},
  author={Redmon, J},
  booktitle={Proceedings of the IEEE conference on computer vision and pattern recognition},
  year={2016}
}

@inproceedings{he2016deep,
  title={Deep residual learning for image recognition},
  author={He, Kaiming and Zhang, Xiangyu and Ren, Shaoqing and Sun, Jian},
  booktitle={Proceedings of the IEEE conference on computer vision and pattern recognition},
  pages={770--778},
  year={2016}
}

@article{cheng2023sam,
  title={Sam-med2d},
  author={Cheng, Junlong and Ye, Jin and Deng, Zhongying and Chen, Jianpin and Li, Tianbin and Wang, Haoyu and Su, Yanzhou and Huang, Ziyan and Chen, Jilong and Jiang, Lei and others},
  journal={arXiv preprint arXiv:2308.16184},
  year={2023}
}

@inproceedings{he2017mask,
  title={Mask r-cnn},
  author={He, Kaiming and Gkioxari, Georgia and Doll{\'a}r, Piotr and Girshick, Ross},
  booktitle={Proceedings of the IEEE international conference on computer vision},
  pages={2961--2969},
  year={2017}
}

@inproceedings{ronneberger2015u,
  title={U-net: Convolutional networks for biomedical image segmentation},
  author={Ronneberger, Olaf and Fischer, Philipp and Brox, Thomas},
  booktitle={Medical image computing and computer-assisted intervention--MICCAI 2015: 18th international conference, Munich, Germany, October 5-9, 2015, proceedings, part III 18},
  pages={234--241},
  year={2015},
  organization={Springer}
}

@inproceedings{cao2022swin,
  title={Swin-unet: Unet-like pure transformer for medical image segmentation},
  author={Cao, Hu and Wang, Yueyue and Chen, Joy and Jiang, Dongsheng and Zhang, Xiaopeng and Tian, Qi and Wang, Manning},
  booktitle={European conference on computer vision},
  pages={205--218},
  year={2022},
  organization={Springer}
}

@inproceedings{milletari2016v,
  title={V-net: Fully convolutional neural networks for volumetric medical image segmentation},
  author={Milletari, Fausto and Navab, Nassir and Ahmadi, Seyed-Ahmad},
  booktitle={2016 fourth international conference on 3D vision (3DV)},
  pages={565--571},
  year={2016},
  organization={Ieee}
}

@misc{abdi2017panoramic,
  author       = {Amir Abdi and Shohreh Kasaei},
  title        = {Panoramic Dental X-rays With Segmented Mandibles},
  year         = {2017},
  howpublished = {\url{https://doi.org/10.17632/hxt48yk462.1}},
  note         = {Mendeley Data, V1},
  doi          = {10.17632/hxt48yk462.1}
}

@article{arsiwala2023machine,
  title={Machine learning in dentistry: a scoping review},
  author={Arsiwala-Scheppach, Lubaina T and Chaurasia, Akhilanand and Mueller, Anne and Krois, Joachim and Schwendicke, Falk},
  journal={Journal of Clinical Medicine},
  volume={12},
  number={3},
  pages={937},
  year={2023},
  publisher={MDPI}
}

@InProceedings{10.1007/978-3-031-88977-6_15,
author="Fu, Kaiwen
and Chang, Chengyuan
and Chen, Jiahui
and Hu, Qinjie",
editor="Wang, Yaqi
and Qian, Dahong
and Wang, Shuai
and Ben-Hamadou, Achraf
and Pujades, Sergi
and Lumetti, Luca
and Grana, Costantino
and Bolelli, Federico",
title="A Self-training Pipeline for Semi-supervised 2D Teeth Instance Segmentation",
booktitle="Supervised and Semi-supervised Multi-structure Segmentation and Landmark Detection in Dental Data",
year="2025",
publisher="Springer Nature Switzerland",
address="Cham",
pages="156--165",
isbn="978-3-031-88977-6"
}

@ARTICLE{10820520,
  author={Xu, Xu and Chen, Junxin and Yin, Jiayue},
  journal={IEEE Journal of Biomedical and Health Informatics}, 
  title={Tooth Instance Segmentation and Disease Detection With Uncertainty-Aware Contrastive Learning and Cross-Scale Attention}, 
  year={2025},
  volume={},
  number={},
  pages={1-12},
  doi={10.1109/JBHI.2024.3525460}}

@article{spector2008implant,
  author    = {Spector, L.},
  title     = {Computer-aided dental implant planning},
  journal   = {Dental Clinics of North America},
  volume    = {52},
  number    = {4},
  pages     = {761--775, vi},
  year      = {2008},
  month     = {October},
  doi       = {10.1016/j.cden.2008.05.004},
  pmid      = {18805228},
  publisher = {Elsevier}
}

@inproceedings{kirillov2023segment,
  title={Segment anything},
  author={Kirillov, Alexander and Mintun, Eric and Ravi, Nikhila and Mao, Hanzi and Rolland, Chloe and Gustafson, Laura and Xiao, Tete and Whitehead, Spencer and Berg, Alexander C and Lo, Wan-Yen and others},
  booktitle={Proceedings of the IEEE/CVF international conference on computer vision},
  pages={4015--4026},
  year={2023}
}

@article{dot2024dentalsegmentator,
  title={DentalSegmentator: robust open source deep learning-based CT and CBCT image segmentation},
  author={Dot, Gauthier and Chaurasia, Akhilanand and Dubois, Guillaume and Savoldelli, Charles and Haghighat, Sara and Azimian, Sarina and Taramsari, Ali Rahbar and Sivaramakrishnan, Gowri and Issa, Julien and Dubey, Abhishek and others},
  journal={Journal of Dentistry},
  volume={147},
  pages={105130},
  year={2024},
  publisher={Elsevier}
}

@article{wang2024trans,
  title={Trans-VNet: Transformer-based tooth semantic segmentation in CBCT images},
  author={Wang, Chen and Yang, Jingyu and Wu, Baoyu and Liu, Ruijun and Yu, Peng},
  journal={Biomedical Signal Processing and Control},
  volume={97},
  pages={106666},
  year={2024},
  publisher={Elsevier}
}

@ARTICLE{9686728,
  author={Cipriano, Marco and Allegretti, Stefano and Bolelli, Federico and Di Bartolomeo, Mattia and Pollastri, Federico and Pellacani, Arrigo and Minafra, Paolo and Anesi, Alexandre and Grana, Costantino},
  journal={IEEE Access}, 
  title={Deep Segmentation of the Mandibular Canal: A New 3D Annotated Dataset of CBCT Volumes}, 
  year={2022},
  volume={10},
  number={},
  pages={11500-11510},
  doi={10.1109/ACCESS.2022.3144840}}

@ARTICLE{9083982,
  author={Chen, Yanlin and Du, Haiyan and Yun, Zhaoqiang and Yang, Shuo and Dai, Zhenhui and Zhong, Liming and Feng, Qianjin and Yang, Wei},
  journal={IEEE Access}, 
  title={Automatic Segmentation of Individual Tooth in Dental CBCT Images From Tooth Surface Map by a Multi-Task FCN}, 
  year={2020},
  volume={8},
  number={},
  pages={97296-97309},
  doi={10.1109/ACCESS.2020.2991799}}

@misc{
 vzrad2_dataset,
title = { vzrad2 Dataset },
type = { Open Source Dataset },
author = { Arshs Workspace Radio },
howpublished = { \url{ https://universe.roboflow.com/arshs-workspace-radio/vzrad2 } },
url = { https://universe.roboflow.com/arshs-workspace-radio/vzrad2 },
journal = { Roboflow Universe },
publisher = { Roboflow },
year = { 2024 },
month = { sep },
}

@misc{hamamci2023dentexabnormaltoothdetection,
      title={DENTEX: An Abnormal Tooth Detection with Dental Enumeration and Diagnosis Benchmark for Panoramic X-rays}, 
      author={Ibrahim Ethem Hamamci and Sezgin Er and Enis Simsar and Atif Emre Yuksel and Sadullah Gultekin and Serife Damla Ozdemir and Kaiyuan Yang and Hongwei Bran Li and Sarthak Pati and Bernd Stadlinger and Albert Mehl and Mustafa Gundogar and Bjoern Menze},
      year={2023},
      eprint={2305.19112},
      archivePrefix={arXiv},
      primaryClass={cs.CV},
      url={https://arxiv.org/abs/2305.19112}, 
}

@inproceedings{lee2013pseudo,
  title={Pseudo-label: The simple and efficient semi-supervised learning method for deep neural networks},
  author={Lee, Dong-Hyun and others},
  booktitle={Workshop on challenges in representation learning, ICML},
  volume={3},
  pages={896},
  year={2013},
  organization={Atlanta}
}

@article{sohn2020fixmatch,
  title={Fixmatch: Simplifying semi-supervised learning with consistency and confidence},
  author={Sohn, Kihyuk and Berthelot, David and Carlini, Nicholas and Zhang, Zizhao and Zhang, Han and Raffel, Colin A and Cubuk, Ekin Dogus and Kurakin, Alexey and Li, Chun-Liang},
  journal={Advances in neural information processing systems},
  volume={33},
  pages={596--608},
  year={2020}
}

@article{tarvainen2017mean,
  title={Mean teachers are better role models: Weight-averaged consistency targets improve semi-supervised deep learning results},
  author={Tarvainen, Antti and Valpola, Harri},
  journal={Advances in neural information processing systems},
  volume={30},
  year={2017}
}

@article{berthelot2019mixmatch,
  title={Mixmatch: A holistic approach to semi-supervised learning},
  author={Berthelot, David and Carlini, Nicholas and Goodfellow, Ian and Papernot, Nicolas and Oliver, Avital and Raffel, Colin A},
  journal={Advances in neural information processing systems},
  volume={32},
  year={2019}
}

@inproceedings{chen2020simple,
  title={A simple framework for contrastive learning of visual representations},
  author={Chen, Ting and Kornblith, Simon and Norouzi, Mohammad and Hinton, Geoffrey},
  booktitle={International conference on machine learning},
  pages={1597--1607},
  year={2020},
  organization={PmLR}
}

@article{cheplygina2019not,
  title={Not-so-supervised: a survey of semi-supervised, multi-instance, and transfer learning in medical image analysis},
  author={Cheplygina, Veronika and De Bruijne, Marleen and Pluim, Josien PW},
  journal={Medical image analysis},
  volume={54},
  pages={280--296},
  year={2019},
  publisher={Elsevier}
}

@article{tajbakhsh2020embracing,
  title={Embracing imperfect datasets: A review of deep learning solutions for medical image segmentation},
  author={Tajbakhsh, Nima and Jeyaseelan, Laura and Li, Qian and Chiang, Jeffrey N and Wu, Zhihao and Ding, Xiaowei},
  journal={Medical image analysis},
  volume={63},
  pages={101693},
  year={2020},
  publisher={Elsevier}
}

@article{isensee2021nnu,
  title={nnU-Net: a self-configuring method for deep learning-based biomedical image segmentation},
  author={Isensee, Fabian and Jaeger, Paul F and Kohl, Simon AA and Petersen, Jens and Maier-Hein, Klaus H},
  journal={Nature methods},
  volume={18},
  number={2},
  pages={203--211},
  year={2021},
  publisher={Nature Publishing Group}
}

\end{document}